\documentclass[a4paper,twoside,british,useAMS,usenatbib]{mn2e}
\usepackage{lmodern}

\usepackage[T1]{fontenc}
\usepackage[latin9]{inputenc}
\usepackage{babel}
\usepackage{textcomp}
\usepackage{amsmath}
\usepackage{amssymb}
\usepackage{graphicx}
\usepackage{esint}
\usepackage[authoryear]{natbib}
\usepackage[unicode=true,
 bookmarks=true,bookmarksnumbered=false,bookmarksopen=false,
 breaklinks=false,pdfborder={0 0 0},backref=false,colorlinks=false]
 {hyperref}
\hypersetup{
 pdfauthor={Lee S. Kelvin},
 pdfsubject={Extragalactic Astronomy}}

\makeatletter

\pdfpageheight\paperheight
\pdfpagewidth\paperwidth

\newcommand{\noun}[1]{\textsc{#1}}
\newcommand{\lyxmathsym}[1]{\ifmmode\begingroup\def\b@ld{bold}
  \text{\ifx\math@version\b@ld\bfseries\fi#1}\endgroup\else#1\fi}

\providecommand{\tabularnewline}{\\}

\renewcommand\[{\begin{equation}}
\renewcommand\]{\end{equation}}
\usepackage{aas_macros}

\makeatother

\begin{document}
\title[GAMA: Sérsic LFs and the CSED by Hubble type]{Galaxy And Mass Assembly (GAMA): $ugrizYJHK$ Sérsic luminosity functions and the cosmic spectral energy distribution by Hubble type}

\author[L.~S.~Kelvin et al.]{
\parbox{\textwidth}{
\raggedright
Lee~S.~Kelvin,$^{1,2,3}$
Simon~P.~Driver,$^{2,3}$
Aaron~S.~G.~Robotham,$^{2,3}$
Alister~W.~Graham,$^{4}$
Steven~Phillipps,$^{5}$
Nicola~K.~Agius,$^{6}$
Mehmet~Alpaslan,$^{2,3}$
Ivan~Baldry,$^{7}$
Steven~P.~Bamford,$^{8}$
Joss~Bland-Hawthorn,$^{9}$
Sarah~Brough,$^{10}$
Michael~J.~I.~Brown,$^{11}$
Matthew~Colless,$^{12}$
Christopher~J.~Conselice,$^{8}$
Andrew~M.~Hopkins,$^{10}$
Jochen~Liske,$^{13}$
Jon~Loveday,$^{14}$
Peder~Norberg,$^{15}$
Kevin~A.~Pimbblet,$^{11,16,17}$
Cristina~C.~Popescu,$^{6}$
Matthew~Prescott,$^{18}$
Edward~N.~Taylor,$^{19}$
and Richard~J.~Tuffs$^{20}$
}\vspace{0.4cm}\\
\parbox{\textwidth}{
$^{1}$Institut f\"{u}r Astro- und Teilchenphysik, Universit\"{a}t Innsbruck, Technikerstra{\ss}e 25, 6020 Innsbruck, Austria\\
$^{2}$School of Physics and Astronomy, University of St Andrews, North Haugh, St Andrews, Fife, KY16 9SS, UK\\
$^{3}$International Centre for Radio Astronomy Research, 7 Fairway, The University of Western Australia, Crawley, Perth, Western Australia 6009, Australia\\
$^{4}$Centre for Astrophysics and Supercomputing, Swinburne University of Technology, Hawthorn, Victoria 3122, Australia\\
$^{5}$Astrophysics Group, H.H. Wills Physics Laboratory, University of Bristol, Tyndall Avenue, Bristol BS8 1TL, UK\\
$^{6}$Jeremiah Horrocks Institute, University of Central Lancashire, Preston PR1 2HE, UK\\
$^{7}$Astrophysics Research Institute, Liverpool John Moores University, Twelve Quays House, Egerton Wharf, Birkenhead CH41 1LD, UK\\
$^{8}$School of Physics and Astronomy, The University of Nottingham, University Park, Nottingham, NG7 2RD, UK\\
$^{9}$Sydney Institute for Astronomy, School of Physics A28, University of Sydney, NSW 2006, Australia\\
$^{10}$Australian Astronomical Observatory, PO Box 915, North Ryde, NSW 1670, Australia\\
$^{11}$School of Physics, Monash University, Clayton, VIC 3800, Australia\\
$^{12}$Research School of Astronomy and Astrophysics, The Australian National University, Canberra, ACT 2611, Australia\\
$^{13}$European Southern Observatory, Karl-Schwarzschild-Str. 2, 85748 Garching, Germany\\
$^{14}$Astronomy Centre, University of Sussex, Falmer, Brighton BN1 9QH, UK\\
$^{15}$Institute for Computational Cosmology, Department of Physics, Durham University, South Road, Durham DH1 3LE, UK\\
$^{16}$Department of Physics, University of Oxford, Denys Wilkinson Building, Keble Road, Oxford OX1 3RH, UK\\
$^{17}$Department of Physics and Mathematics, University of Hull, Cottingham Road, Hull, HU6 7RX, UK\\
$^{18}$Astrophysics Group, Department of Physics, University of the Western Cape, 7535 Bellville, Cape Town, South Africa\\
$^{19}$School of Physics, the University of Melbourne, Parkville, VIC 3010, Australia\\
$^{20}$Max Planck Institut f\"{u}r Kernphysik, Saupfercheckweg 1, D-69117 Heidelberg, Germany\\
}
\vspace{-0.75cm}
}

\date{Draft}

\pagerange{\pageref{firstpage}--\pageref{lastpage}} \pubyear{2013}

\maketitle

\label{firstpage}
\begin{abstract}
We report the morphological classification of $3727$ galaxies from
the Galaxy and Mass Assembly survey with $M_{r}<-17.4$ mag and in
the redshift range $0.025<z<0.06$ ($2.1\times10^{5}$ Mpc$^{3}$)
into E, S0-Sa, SB0-SBa, Sab-Scd, SBab-SBcd, Sd-Irr and little blue
spheroid classes. Approximately $70$\% of galaxies in our sample
are disk dominated systems, with the remaining $\sim30$\% spheroid
dominated. We establish the robustness of our classifications, and
use them to derive morphological-type luminosity functions and luminosity
densities in the $ugrizYJHK$ passbands, improving on prior studies
that split by global colour or light profile shape alone. We find
that the total galaxy luminosity function is best described by a double-Schechter
function while the constituent morphological-type luminosity functions
are well described by a single-Schechter function.

These data are also used to derive the star-formation rate densities
for each Hubble class, and the attenuated and unattenuated (corrected
for dust) cosmic spectral energy distributions, i.e., the instantaneous
energy production budget. While the observed optical/near-IR energy
budget is dominated $58$:$42$ by galaxies with a significant spheroidal
component, the actual energy production rate is reversed, i.e., the
combined disk dominated populations generate $\sim1.3\times$ as much
energy as the spheroid dominated populations. On the grandest scale,
this implies that chemical evolution in the local Universe is currently
confined to mid-type spiral classes like our Milky Way.
\end{abstract}
\begin{keywords}
galaxies: luminosity function, mass function -- galaxies: fundamental parameters -- galaxies: elliptical and lenticular, cD -- galaxies: spiral
\vspace{-0.5cm}
\end{keywords}

\section{Introduction}

\label{sec:intro}In his seminal 1926 paper `Extra-galactic Nebulae',
Edwin Hubble established a framework for the morphological classification
of galaxies which remains in use essentially unchanged to the present
day. From a sample of 400 galaxies, and perhaps drawing inspiration
from \citet{Jeans1919} and \citet{Reynolds1920}, Hubble defined
three main sub-groups; Elliptical, Spiral and Lenticular \citep{Hubble1926,Hubble1936a}.
Elliptical early-type galaxies typically show no additional structure
beyond a smooth radial light profile. Conversely, late-type spiral
galaxies consist of a central spheroidal bulge surrounded by a flattened
extended disk containing spiral arm features, and perhaps with the
presence of a bar. Lenticular galaxies fall somewhere in-between,
with the familiar late-type bulge and disk features present, potentially
with the addition of a bar, and yet the noticeable absence of spiral
arm structure. 

Many additions have been suggested to Hubble's classification scheme,
in order to account for, e.g.: the presence of rings \citep{Sandage1961};
transition lenticular galaxies \citep{Holmberg1958}; bulge-less Sd-type
disk galaxies \citep{Shapley1940}; the `boxy' and `disky' isophotes
of early-type galaxies \citep{Carter1978,Davies1983a,Carter1987,Kormendy1996};
the large variation in lenticular bulge-to-disk ratios \citet{vandenBergh1976,Laurikainen2010,Cappellari2011a,Kormendy2012},
and; the presence of dwarf galaxies \citep{Shapley1938,Sandage1984}.
These additions each provide important information to morphological
classification schema, adding additional resolution to each classification
element. 

Moreover, while morphology is intrinsically linked to the star formation
rate of the galaxy, it was shown by \citet{Dressler1980} that the
distribution of morphological types varies as a function of local
galaxy density: the morphology-density relation. Many possible explanations
for this exist in the literature, including four key effects: strangulation
\citep{Larson1980,Kauffmann1993,Diaferio2001}, harassment \citep{Moore1996},
ram pressure stripping \citep{Gunn1972} and minor-merging or tidal
interactions \citep{Park2008}. Each of these mechanisms in some way
affects the star formation rate of the interacting system, shutting
off star formation for galaxies in over-dense regions and consequently
causing a change in colour and ultimately morphology. Possibly more
fundamental than the relation between morphology and environment is
the connection between galaxy structure (i.e.; bulge, disk, bar) and
its host galaxy's stellar mass \citep{vanderWel2008,Bamford2009,Nair2010,Wilman2012,Pimbblet2012}.

The logical basis for defining these morphological groupings remains
a visual one, and so becomes increasingly time consuming in an era
of large scale observational astronomy. Despite this, the scientific
worth of morphological classification remains extremely high. The
morphological class of a galaxy is a tracer of its evolutionary history,
with merging events believed to be the primary cause of the transition
of late-type spirals into early-type ellipticals (e.g., \citealp{Park2008}).

In this paper we provide morphological classifications for a local
($0.025<z<0.06$) volume limited sample of $3,727$ galaxies brighter
than $M_{r,S\acute{e}rsic}=-17.4$ mag taken from the Galaxy and Mass
Assembly survey (GAMA; \citealp{Driver2009}). Using these classifications,
we measure the global and constituent morphological-type luminosity
functions in optical $ugriz$%
\footnote{These passbands have effective wavelength midpoints of $354$, $475$,
$622$, $763$ and $905$ nm, respectively.%
} and near-infrared $YJHK$%
\footnote{These passbands have effective wavelength midpoints of $1031$, $1248$,
$1631$ and $2201$ nm, respectively.%
} passbands.

This paper is structured as follows. We define our sample and postage
stamp cutout creation in Section \ref{sec:data}. This sample is morphologically
classified by eye by three independent observers, described in Section
\ref{sec:classification}. We explore the trends with morphology against
complementary global galaxy measurements such as colour, stellar mass
and Sérsic index in Section \ref{sec:trends}. We present the global
and the individual morphological-type luminosity functions for all
nine passbands in Section \ref{sec:lumfuncs}, and discuss the division
of the cosmic spectral energy distribution by morphology, both with
and without suitable corrections to account for dust attenuation,
in Section \ref{sec:csed}. A standard cosmology of ($H_{0}$, $\Omega_{m}$,
$\Omega_{\Lambda}$)$=$($70$ km s$^{-1}$ Mpc$^{-1}$, $0.3$, $0.7$)
is assumed throughout this paper.

\section{Data}

\label{sec:data}Our data are taken from the GAMA survey \citep{Driver2009,Driver2011},
specifically GAMA phase $1$, known as GAMA-I. GAMA is a combined
spectroscopic and multi-wavelength imaging programme designed to study
spatial structure in the nearby ($z<0.25$) Universe on kpc to Mpc
scales (see \citealt{Driver2009} for an overview). The survey, after
completion of phase $1$, consists of three regions of sky each of
$4$ deg (Dec) $\times12$ deg (RA), close to the equatorial region,
at approximately 9$^{h}$ ($135$ deg; G09), 12$^{h}$ ($180$ deg;
G12) and 14.5$^{h}$ ($217.5$ deg; G15). The three regions were selected
to enable accurate characterisation of the large scale structure over
a range of redshifts and with regard to practical observing considerations
and constraints. They lie within areas of sky surveyed by both the
Sloan Digital Sky Survey (SDSS; \citealt{York2000,Abazajian2009})
as part of its Main Survey, and by the United Kingdom Infrared Telescope
(UKIRT) as part of the UKIRT Infrared Deep Sky Survey (UKIDSS) Large
Area Survey (UKIDSS-LAS; \citealp{Lawrence2007}). These data provide
moderate depth and resolution imaging in $ugrizYJHK$ suitable for
analysis of nearby galaxies. GAMA imaging data presented in this paper
is constructed from reprocessed SDSS and UKIDSS-LAS imaging data,
rescaled to a common pixel scale of $0.339''$/pixel and to a common
zero point magnitude of $30$ mag arcsec$^{-2}$. Further details
on the GAMA imaging pipeline may be found in \citet{Hill2011}. The
accompanying spectroscopic input catalogue was derived from the SDSS
PHOTO parameter \citep{Stoughton2002} as described in \citet{Baldry2010}.
The GAMA spectroscopic programme \citep{Robotham2010} commenced in
2008 using 2dF+AAOmega on the Anglo-Australian Telescope to obtain
distance information (redshifts) for all galaxies brighter than $r<19.8$
mag. The survey is $\sim99$ per cent complete to $r<19.4$ mag in
G09 and G15 and $r<19.8$ mag in G12, with a median redshift of $z\sim0.2$.
Full details of the GAMA Phase I (GAMA-I) spectroscopic programme,
key survey diagnostics, and the GAMA public and team databases are
given in \citet{Driver2011} and \citet{Hopkins2013}.

\subsection{Luminosity Limits}

\label{sub:absmaglimits}

\subsubsection{Absolute Sérsic Magnitudes}

\label{sub:abssersicmags}Although the GAMA survey limits mentioned
above are defined using SDSS Petrosian photometry, our preferred measure
of total magnitudes are those derived from truncated single-Sérsic
fits to the data (see \citealp{Kelvin2012}). Initially a generalisation
of the \citet{deVaucouleurs1948} $r^{1/4}$ model for describing
the radial light profiles of early-type galaxies, the \citet{Sersic1963,Sersic1968}
$r^{1/n}$ model, subsequently reviewed in \citet{Graham2005a}, has
become a standard tool for quantifying galaxies across a wide range
of morphologies, both early- and late-type. The Sérsic equation provides
the intensity $I$ at a given radius $r$ as given by
\begin{equation}
I\left(r\right)=I_{e}\exp\left[-b_{n}\left(\left(\frac{r}{r_{e}}\right)^{1/n}-1\right)\right]\label{eq:sersic}
\end{equation}
where $I_{e}$ is the intensity at the effective radius $r_{e}$,
the radius containing half of the projected total light, and $n$
is the Sérsic index which determines the shape of the light profile.
The value of $b_{n}$ is a function of Sérsic index, as defined in
\citet{Ciotti1991}, and is such that $\Gamma(2n)=2\gamma(2n,b_{n})$,
where $\Gamma$ and $\gamma$ represent the complete and incomplete
gamma functions respectively%
\footnote{$b_{n}$ can trivially be calculated using the programming language
R using the relation $b_{n}=\mathrm{qgamma}(0.5,2n)$, where $\mathrm{qgamma}$
is the quantile function for the Gamma distribution. For the range
$0.5<n<10$, \citet{Capaccioli1989} approximates the value of $b_{n}$
using the relation $b_{n}=1.9992n\lyxmathsym{\textminus}0.3271$.%
}. Single-Sérsic model fits have been shown to provide a good description
of galaxy light profiles as faint as $B\sim28$ mag/arcsec$^{2}$
(\citealp{Caon1990,Caon1993,Caon1994}). Therefore, Sérsic modelling
allows for us to account for the missing flux in the wings of high
central-concentration galaxies, side-stepping the well documented
problems with both Petrosian and Kron photometry (see, e.g., \citealp{Graham2005a,Graham2005b}).
We elect to truncate our Sérsic magnitudes at $10$ multiples of the
half-light radius ($10$ $r_{e}$). This is to avoid extrapolation
of flux into regimes below the limiting isophote for which we are
uncertain of the true light profile of the galaxy: consequently not
parametrising our ignorance. For further discussion of Sérsic photometry
and truncation, see \citet{Kelvin2012}.

For a given band $x$, absolute Sérsic magnitudes $M_{x}$ are derived
using the standard relation
\begin{equation}
M_{x}=m_{x}-\left(5\log_{10}\mathrm{D_{L}}+25\right)-k_{x}-e_{x}-A_{x}\label{eq:abmag}
\end{equation}
where $m_{x}$ denotes the apparent magnitude (in this case, truncated
Sérsic), $\mathrm{D_{L}}$ is the luminosity distance of the galaxy
in megaparsecs (where $\mathrm{D_{L}}$ is related to the angular
diameter distance, $\mathrm{D_{A}}$, using the relation $\mathrm{D_{L}}=(1+z)^{2}\mathrm{D_{A}}$),
$k_{x}$ is the applied $k$-correction for band $x$, $e_{x}$ is
the evolutionary correction and $A_{x}$ is the Milky Way dust attenuation
correction. We obtain appropriate $k$-corrections from version $8$
of the GAMA-I stellar masses catalogue \emph{(StellarMassesv08}; see
\citealp{Taylor2011}). One would expect minimal evolutionary effects
over the narrow redshift range of this sample (see, e.g., \citealp{Prescott2009}),
and so we do not apply any $e$-corrections to these data. We apply
the Milky Way dust attenuation correction as given in Table $22$
of \citet{Stoughton2002}, with UKIDSS values determined by matching
UKIDSS database values from the WFCAM Science Archive to the SDSS
extinction in the $r$ band. Further details on this procedure may
be found in Liske et al. (in prep.).

\subsubsection{Absolute Sérsic Magnitude Limit in the r Band}

In order to avoid the many incompleteness issues affecting the dwarf
systems (both Malmquist and surface-brightness bias), in this study
we focus on the more luminous systems by removing those galaxies with
an absolute Sérsic magnitude in the $r$ band fainter than $M_{r}=-17.4$
mag. This value is determined based on the faintest magnitude down
to which our GAMA sample would be complete at our upper redshift limit
of $z=0.06$ ($m_{r}\sim19.4$ mag; see Appendix \ref{app:redshiftlimits}
and Figure \ref{fig:gamanear}). Sérsic and structural measurements
are taken from version $7$ of the GAMA-I Sérsic photometry catalogue
(\emph{SersicPhotometryv07}; \citealp{Kelvin2012}). To summarise
this study; a 2-dimensional single-Sérsic model is fit to each galaxy
in our sample using the SIGMA galaxy fitting pipeline. SIGMA is a
wrapper around several contemporary commonplace astronomy tools including
Source Extractor \citep{Bertin1996}, PSF Extractor \citep{Bertin2011}
and GALFIT \citep{Peng2010a}.

\subsection{Sample Definition}

\label{sub:sample}Using the latest version (version $16$) of the
GAMA-I tiling catalogue%
\footnote{All GAMA catalogues are available through the GAMA database, available
online at http://www.gama-survey.org/.%
} (\emph{TilingCatv16}, see \citealp{Baldry2010}) we define a volume
limited sample of $3,727$ galaxy-like ($\mathrm{SURVEY\_CLASS}\geq2$)
objects whose local flow-corrected redshifts $z$ lie in the range
$0.025<z<0.06$ (see Appendix \ref{app:redshiftlimits}) with an associated
normalised redshift quality $nQ>2$ (i.e., good for science%
\footnote{GAMA spectroscopic redshifts are assigned a quality from $0$ to $4$,
where $0$ is a corrupted/bad spectrum and therefore a meaningless
associated redshift, and $4$ is a high-quality redshift with a high
degree of certainty. Typically, we advocate using $Q>2$ for scientific
analyses.%
}), an extinction corrected $r$ band SDSS Petrosian magnitude of $r<19.4$
mag (ensuring a consistent depth across all three GAMA regions) and
an absolute truncated Sérsic magnitude in the $r$-band of $M_{r}<-17.4$
mag. This luminosity and volume-limited sample of $3,727$ galaxies
is referred to as GAMAnear.

Our redshift limits give this sample a volume of $2.1\times10^{5}$
Mpc$^{3}$. Note that redshifts have been matched from version $7$
of the local flow corrected redshift catalogue (\emph{DistancesFramesv07}),
itself based on data from version $8$ of the GAMA-I spectroscopic
catalogue (\emph{SpecObjv08}). These redshifts are Milky Way centric,
but local velocity field effects have been removed. Matching to the
GAMA galaxy group catalogue (G3C; \citealp{Robotham2011b}), we find
that just under half ($1797$, $\sim48\%$) of our galaxies lie in
identified groups with two or more members, with $672$ galaxies ($\sim18\%$)
in groups with a richness greater than $5$, i.e.; our sample is predominantly
field dominated.

\begin{figure}
\includegraphics[width=1\columnwidth]{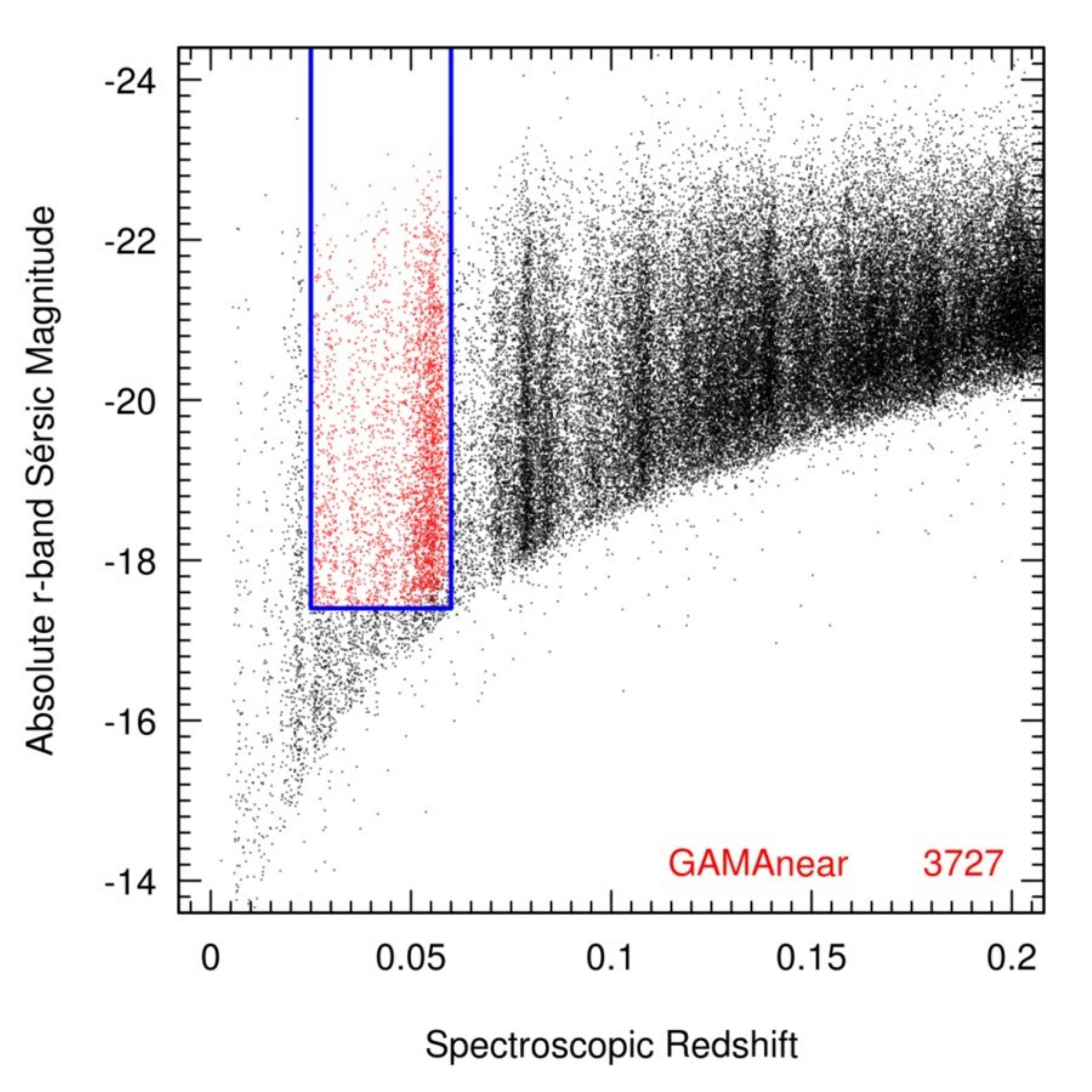}

\caption{\label{fig:gamanear}Absolute Sérsic magnitude ($r$ band) as a function
of local flow corrected spectroscopic redshift for galaxies within
the GAMA survey. Red data points inside the blue box represent the
$3,727$ galaxies in the GAMAnear sample. Local flow-corrected spectroscopic
redshifts are taken from version $7$ of the local flow corrected
redshift catalogue (\emph{DistancesFramesv07}), and absolute Sérsic
magnitudes from version $7$ of the Sérsic photometry catalogue (\emph{SersicCatAllv07};
\citealp{Kelvin2012}).}
\end{figure}

Figure \ref{fig:gamanear} shows absolute Sérsic magnitude ($r$ band)
as a function of local flow corrected spectroscopic redshift for the
full GAMA dataset (black points). Red data points inside the blue
box represent the $3,727$ galaxies in the GAMAnear sample. A summary
of all GAMA data products used to define these samples are shown in
Table \ref{tab:gamadata}.

\setlength{\tabcolsep}{2pt}

\begin{table*}
\begin{tabular}{l|l|l|l|l}
DMU & Version & Catalogue & Paper & Summary of data products used in this study\tabularnewline
\hline 
InputCat & 16 & TilingCat & \citet{Baldry2010} & Target information (RA, Dec, SDSS Petrosian magnitude, Survey Class)\tabularnewline
InputCat & 16 & InputCatA & \citet{Baldry2010} & Extinction corrections\tabularnewline
LocalFlowCorrection & 7 & DistancesFrames & \citet{Baldry2012} & Local flow-corrected spectroscopic redshifts, redshift quality flags\tabularnewline
StellarMasses & 8 & StellarMasses & \citet{Taylor2011} & Global galaxy colours, $k$-corrections (priv. comm.)\tabularnewline
SersicPhotometry & 7 & SersicCatAll & \citet{Kelvin2012} & Sérsic photometry, Structural measurements\tabularnewline
\hline 
\end{tabular}

\caption{\label{tab:gamadata}A summary of the GAMA data products that have
been collated for this study. GAMA catalogues and their associated
data products are grouped into Data Management Units (DMUs), which
are also listed here for reference.}
\end{table*}

\setlength{\tabcolsep}{6pt}

\subsection{Magnitude Limits in Additional Passbands}

The $r$ band absolute magnitude limit for our volume limited GAMAnear
sample ($M_{r}=-17.4$ mag) introduces a colour-dependent limit across
the remaining eight passbands in use from the SDSS and UKIDSS. This
variable limit has the potential to introduce incompleteness bias
when analysing data at other wavelengths, and so (following \citealp{Driver2012}),
we define additional limits down to which the sample remains complete
and unbiased as a function of colour for each passband. 

\begin{figure}
\includegraphics[width=1\columnwidth]{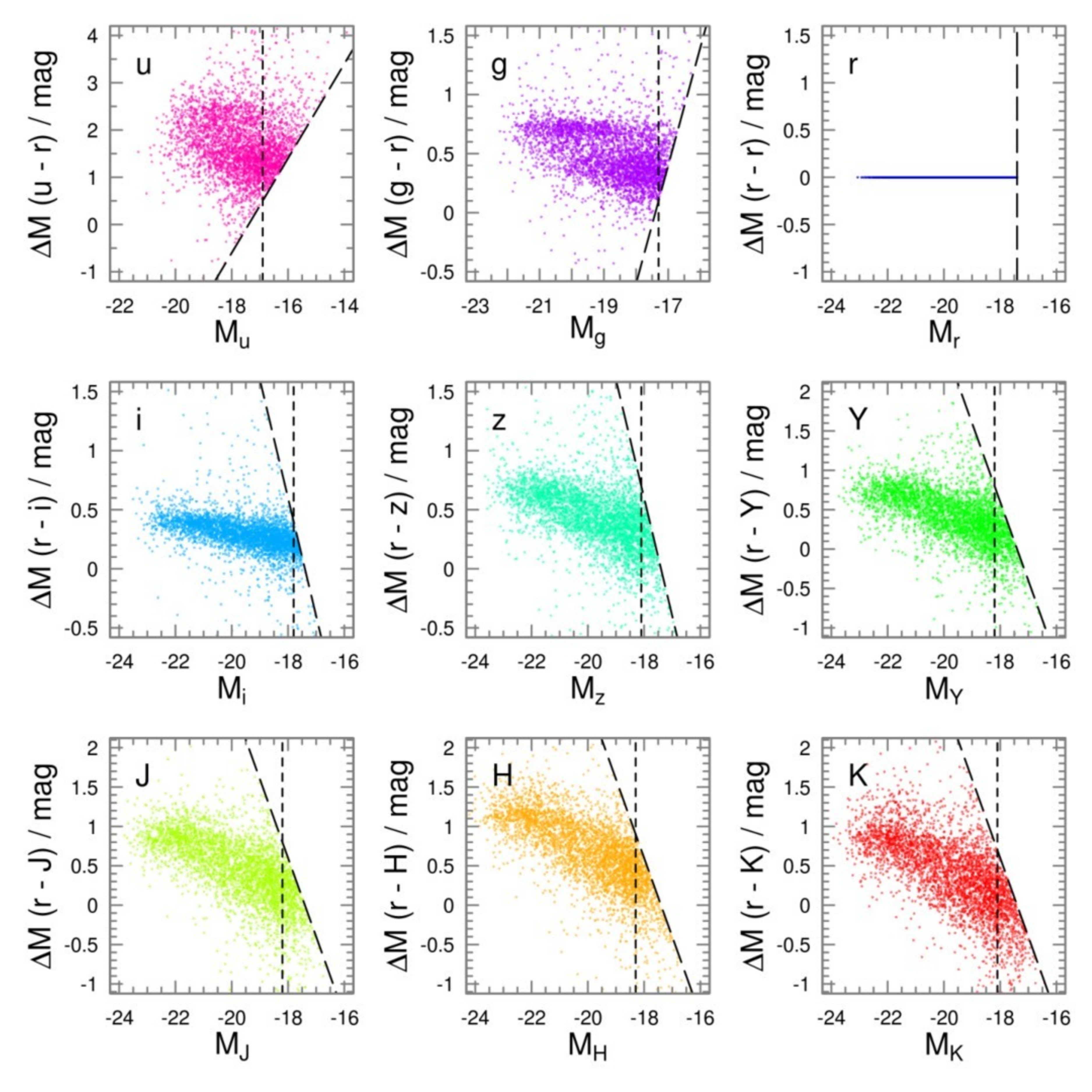}

\caption{\label{fig:colouroffsets}Colour-Magnitude diagrams for all galaxies
in the GAMAnear sample across all nine bands. These values are derived
from absolute Sérsic magnitudes truncated at $10$ $r_{e}$ with k-corrections
and Galactic dust corrections applied. Long-dashed lines represent
the volume-limited sample limit of $M_{r}=-17.4$ mag. Short-dashed
lines represent the absolute magnitude at which the main body of data
intersects the long-dashed line, and shows to what magnitude limit
this sample is complete down to for that wavelength. These limits
are listed in Table \ref{tab:faintlimits}.}
\end{figure}

The colour-magnitude diagrams in Figure \ref{fig:colouroffsets} show
the relation between colour and absolute magnitude for galaxies in
the GAMAnear sample. Long dashed lines represent the $r$ band limit
of $M_{r}=-17.4$ mag. One can clearly see the two distinct populations
(i.e., a bimodal distribution in both colour and absolute magnitude)
in the $g$ band data; the blue cloud and red sequence. These two
populations are also evident to a lesser extent at all wavelengths. 

We define the additional faint-end limits visually as the absolute
magnitude in band $x$ (where $x=ugizYJHK$) at which the main body
of the data intersects the $M_{r}=-17.4$ mag line. These passband
limits are listed in Table \ref{tab:faintlimits}, and shown as vertical
short dashed lines in Figure \ref{fig:colouroffsets}. The absolute
magnitude of the Sun in all passbands is also shown in Table \ref{tab:faintlimits},
for reference (values taken from Table $1$, \citealp{Hill2010}).

\begin{table*}
\begin{centering}
\begin{tabular}{c||ccccccccc}
Band & $u$ & $g$ & $r$ & $i$ & $z$ & $Y$ & $J$ & $H$ & $K$\tabularnewline
\hline 
Limit & $-16.9$ & $-17.3$ & $-17.4$ & $-17.8$ & $-18.1$ & $-18.2$ & $-18.2$ & $-18.3$ & $-18.1$\tabularnewline
Number of galaxies below limit & $2841$ & $3445$ & $3727$ & $3536$ & $3342$ & $3195$ & $3196$ & $3236$ & $3197$\tabularnewline
$\mathrm{M_{\odot}}$ & $6.38$ & $5.15$ & $4.71$ & $4.56$ & $4.54$ & $4.52$ & $4.57$ & $4.71$ & $5.19$\tabularnewline
\end{tabular}
\par\end{centering}

\caption{\label{tab:faintlimits}Absolute Sérsic magnitude limits, galaxy number
counts and absolute solar magnitudes. Sérsic limits denote the faint-end
absolute magnitude at which the sample is complete for that band.
Limits are defined as the absolute magnitude at which the main body
of data in the colour-magnitude diagrams of Figure \ref{fig:colouroffsets}
intersect the volume-limited sample faint-end limit of $M_{r}=-17.4$.
The number of galaxies refers to the how many galaxies from the GAMAnear
sample are brighter than the limit in that band. The absolute magnitude
of the Sun in all passbands is also shown, for reference.}
\end{table*}

\section{Visual Classification}

\label{sec:classification}

\subsection{Classification Criteria}

Perhaps the most simplistic and robust means by which a sample of
galaxies may be classified into their appropriate morphological types
is by visual `eyeball' inspection. We create three-colour postage
stamp images for each galaxy in our GAMAnear sample of $3727$ objects
using the \noun{pi} plotting tool, an internal GAMA software product%
\footnote{A web version of this tool exists at the following web address: http://thuban4.st-and.ac.uk/gama/colcutout/gamacutout.php%
}. For our analysis, we opt to take red, green and blue colours from
the UKIDSS $H$ and SDSS $i$ and $g$ bands, respectively.

Eyeball classification occurs in two phases. Phase $1$ postage stamps
depict $20''\times20''$ with the dynamic range of the images scaled
logarithmically and prior decisions made on the lower (black) and
upper (white) cut levels. Phase $2$ postage stamps depict a larger
area of $40''\times40''$ and are scaled using the arctan function.
We found that the arctan function removes the necessity for a harsh
upper or lower cut level. Imposing harsh cuts has the potential to
lead to misclassification as it imposes an apparent physical boundary
in the light profile of a galaxy where none exists. The increased
area of the phase $2$ postage stamps also allows for the galaxy to
be put into context of its local environment, and allows the observer
to see more than the core of nearby extended galaxies.

\begin{figure*}
\begin{centering}
\includegraphics[width=1\textwidth]{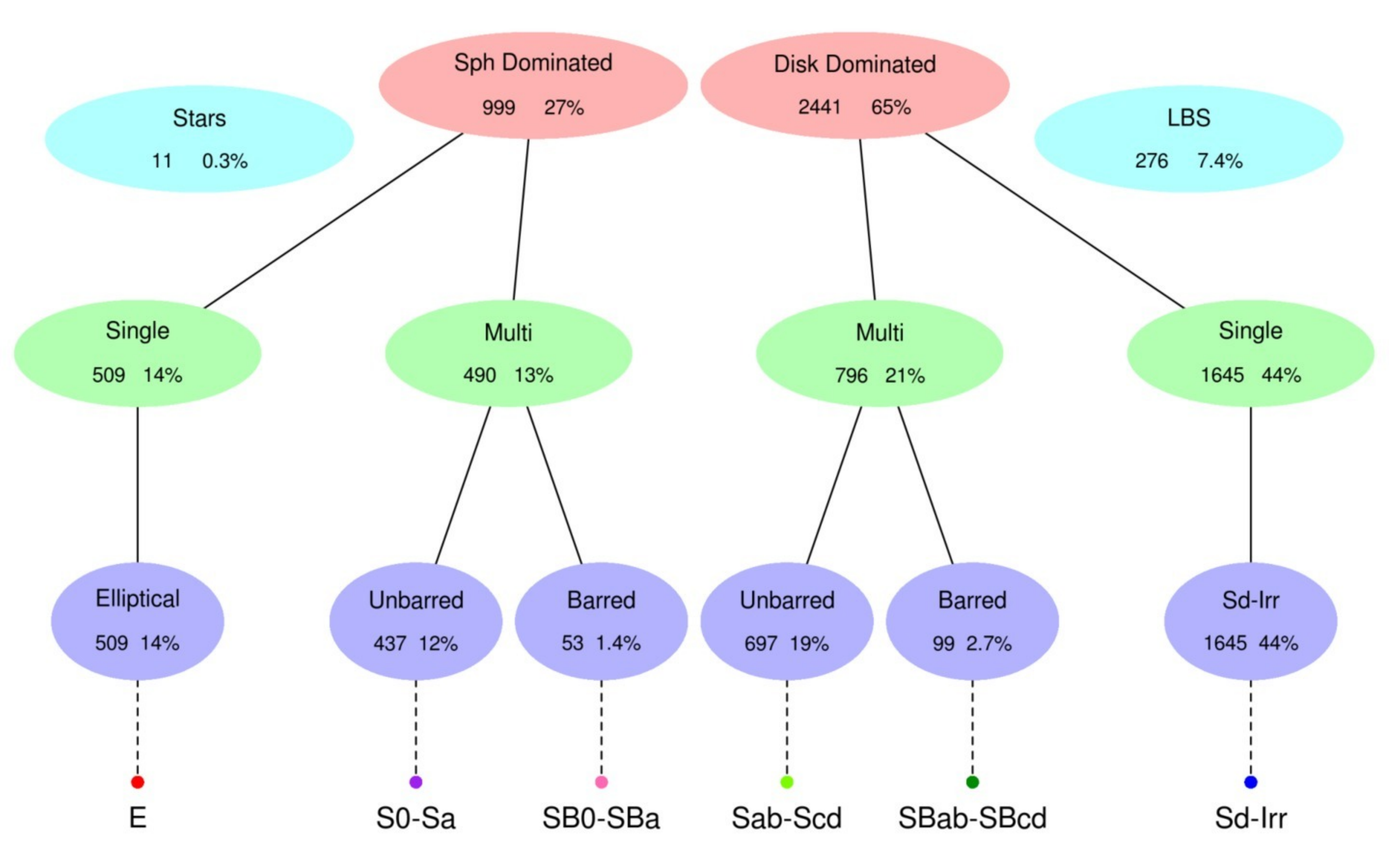}
\par\end{centering}

\caption{\label{fig:numbers}The morphological classification hierarchy used
to filter the volume-limited GAMAnear sample of $3,727$ galaxies
into their appropriate class. In brief: is the galaxy spheroid or
disk dominated?; is the galaxy a single component or a multi-component
system (`single' and `multi')?; and, if the galaxy is a multi-component
system, does it contain a bar (`unbarred' and `barred')? At the top
level, the classes `Stars' and little blue spheroids (LBS) are also
available. See the text for further details. Beneath each label are
the number of galaxies in the master classification bin for that group
and an indication of the fraction of our total sample this group constitutes.
The final morphological type at the bottom of this figure depends
upon the prior decisions made by the classifier.}
\end{figure*}

Classification occurs by assigning the postage stamp of a galaxy into
a specific directory hierarchy. A schematic representation of this
hierarchy is shown in Figure \ref{fig:numbers}, with final `master'
number counts inset for later reference. Postage stamp images are
populated at the top level, and visual classification decisions eventually
filter a galaxy down through this classification tree into its appropriate
morphological class, from E to Sd-Irr, as indicated. The decision
tree is essentially binary at each level (with the exception of stars
and the `Little Blue Spheroid' classes). These levels are Spheroid
Dominated/Disk Dominated, Single/Multi and Barred/Unbarred, and are
discussed here:

\textbf{Spheroid Dominated/Disk Dominated} Galaxies are initially
split into spheroid or disk dominated%
\footnote{Here, the terms `spheroid dominated' and `disk dominated' do not refer
to the spheroidal or disk component dominating the total flux of the
system. As has been shown in \citet{Graham2008b}, rarely does the
spheroid component in a bulge+disk system contribute $>50$\% of the
flux for galaxies later than S0. Rather, we define the term `spheroid
dominated' and `disk dominated' to refer to the visual impact of the
spheroid or the disk on the postage stamp images; a combination of
relative size, apparent surface brightness and 2D light profile.%
}. Colour may be a useful indicator here, however, the apparent gradient
and smoothness of the light profile and the central concentration
are the main discerning factors. 

\textbf{Single/Multi} A question of the total number of distinct structural
components comprising the galaxy. Spheroid-dominated single-component
galaxies are classical elliptical systems, whereas spheroid-dominated
multi-component galaxies are lenticular or early-type spiral systems
(S0-Sa). Disk-dominated single-component galaxies are bulge-less disk
systems or irregulars (Sd-Irr), whereas disk-dominated multi-component
galaxies are late-type spiral systems (Sab-Scd).

\textbf{Barred/Unbarred} The final level of classification determines
whether a multi-component system contains a bar structure. If the
disk is edge-on, and the presence of a bar cannot be verified, then
the galaxy is classified as unbarred. 

At the Spheroid Dominated/Disk Dominated level of classification,
two additional classification options are available: stars and `Little
Blue Spheroids' (LBS). As noted above, this is the only occasion on
which the classification question is not binary. 

\textbf{Stars} If the primary object in the postage stamp depicts
either a foreground star in front of a background galaxy (for which
the associated redshift belongs) or a supernova within a distant galaxy,
it is classified as a star. These objects are removed from the classification
tree at the top level. This class of object is removed in all subsequent
scientific analyses unless specifically mentioned in the text.

\textbf{Little Blue Spheroids (LBS)} Prior to classification it became
apparent that an additional type of galaxy which lies outside the
standard Hubble-Jeans tuning fork exists within our sample. These
galaxies are typically compact, spheroidal and blue, hence their designation
as `Little Blue Spheroids' (LBS from here). The median colour of our
LBS sample is $g-i\sim0.6$ with a median Sérsic index of $n_{r}\sim1.9$
in the $r$ band ($n_{K}\sim1.6$ in the $K$ band) and a median physical
size of $r_{e}\sim1.1$ kpc in the $r$ band ($r_{e}\sim0.9$ kpc
in the $K$ band). LBS-type galaxies may come about via the intermittent
stochastic star formation predicted in low-mass dwarf galaxies by
\citet{Stinson2007}, and have been previously isolated observationally
by \citet{Arp1965,Sandage1984,Guzman1997} and more recently by \citet{Brough2011}
and \citet{Bauer2013}, amongst others. \citet{Brough2011} finds
that these systems are predominantly low-mass and found in low-density
environments, showing similar properties to dwarf irregular galaxies
in the Local Volume. For the purposes of this study, these objects
are removed from the classification tree at the top level.

Three observers; LSK, SPD and ASGR, independently classified the entire
sample of $3,727$ galaxies using both the phase $1$ and phase $2$
postage stamp images. Phase $2$ postage stamp images are initially
placed into their Phase $1$ hierarchy positions as assigned by their
classifier in order to speed up and improve the second round of classification.

\subsection{Classification Results}

\begin{figure}
\begin{centering}
\includegraphics[width=1\columnwidth]{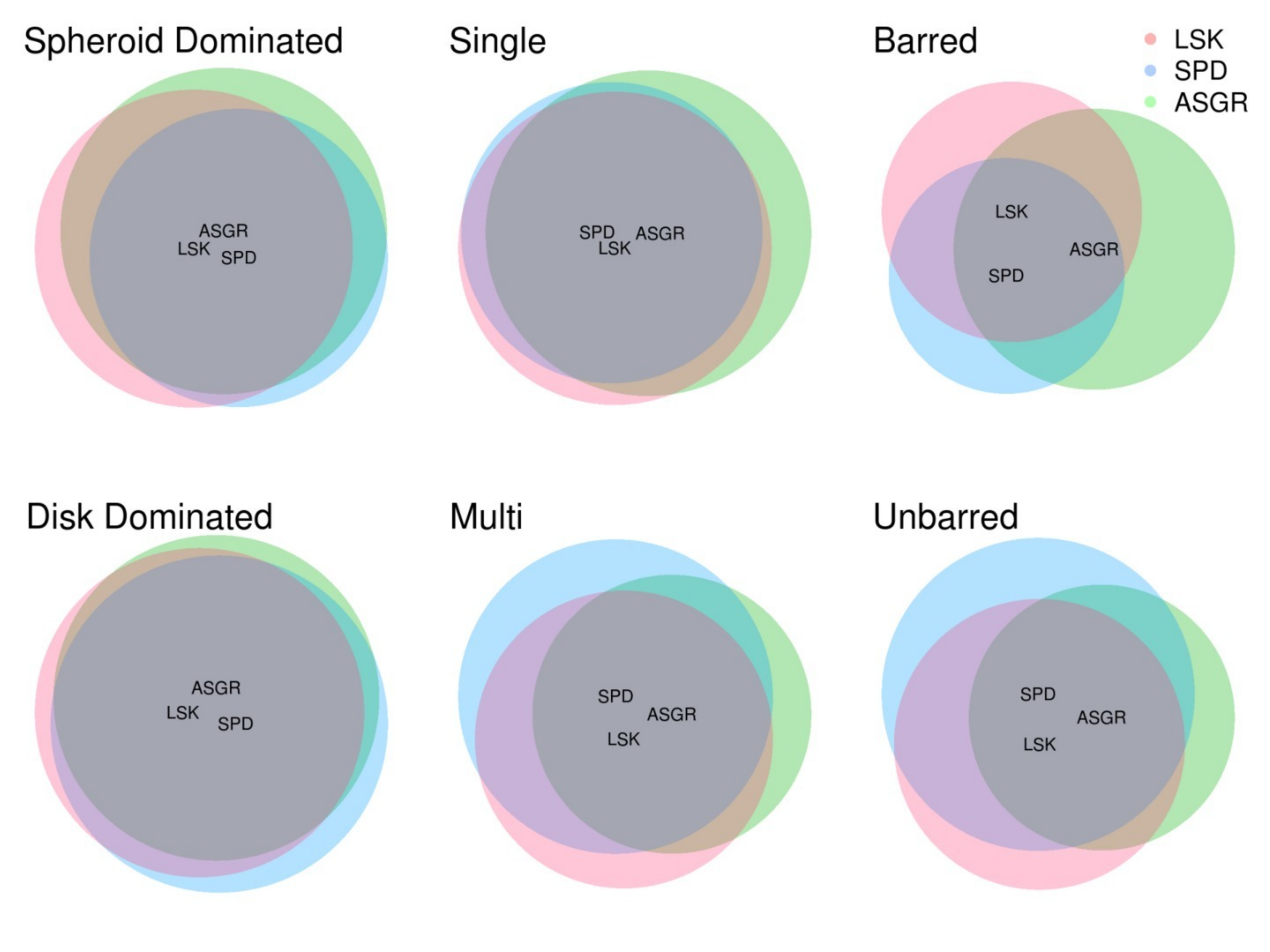}
\par\end{centering}

\caption{\label{fig:euler}Euler-diagrams representing the level of agreement
between the three visual classifiers (LSK, SPD and ASGR) for the six
main decision tree classifications; Spheroid Dominated/Disk Dominated,
Single/Multi and Barred/Unbarred. Objects where any single classifier
classified the system as either `Star' or `LBS' ($451$ in total)
have been removed from this figure, for clarity.}
\end{figure}

A final master classification is assigned based on majority agreement.
In most cases (i.e., Single/Multi and Barred/Unbarred) this requires
the agreement of at least two observers. At the top level (Spheroid
Dominated/Disk Dominated, Stars, LBS), there is a possibility that
all three observers disagree on the classification. In this instance,
a preference is applied to each observer by order of classification
experience (in order: SPD, ASGR, LSK). Should a disagreement arise,
the classification will default to the preferred observer. These weights
also apply at lower levels should a classifier have already been removed
from a classification tree at the top level. At the top level, there
are $56$ such three-way disagreements in our combined GAMAnear sample
of $3,727$ objects ($1.5$\%). In addition, a total of $451$ objects
($12.1$\%) were classified as either `Star' or `LBS' by at least
one observer. A visual representation of the level of agreement between
classifiers on the three standard questions (Spheroid Dominated/Disk
Dominated, Single/Multi, Barred/Unbarred) is shown in Figure \ref{fig:euler}.

Generally there is good agreement between observers, however; all
three observers show a noticeable disagreement on whether a system
hosts a bar, which may explain the relatively low bar fraction in
our galaxy sample. For our $8$ classification bins we find the following
$3$-way agreement fractions: Spheroid Dominated: $19.2\%$ ($714$);
Disk Dominated: $56.5\%$ ($2107$); Single: $41.9\%$ ($1563$);
Multi: $23.5\%$ ($877$); Barred: $2.1\%$ ($77$); Unbarred: $17.7\%$
($659$); Stars: $0.08\%$ ($3$); LBS: $2.5\%$ ($95$).

On combining these classification results using the method outlined
above, just under half of our sample, $44.1$\% ($1,645$), is visually
classified as Sd-Irr type, with elliptical galaxies accounting for
$13.7$\% ($509$) of the sample. Spheroid-dominated multi-component
systems account for $13.1$\% ($490$) of the sample, of which $10.8$\%
($53$) are visually barred. Disk-dominated multi-component systems
account for $21.4$\% ($796$) of the sample, of which $12.4$\% ($99$)
are visually barred. Additionally, $0.3$\% ($11$) of our sample
are classified as `Stars', and $7.4$\% ($276$) as `Little Blue Spheroids'.
These classifications shall be used throughout the remainder of this
study.

\begin{figure*}
\includegraphics[width=1\textwidth]{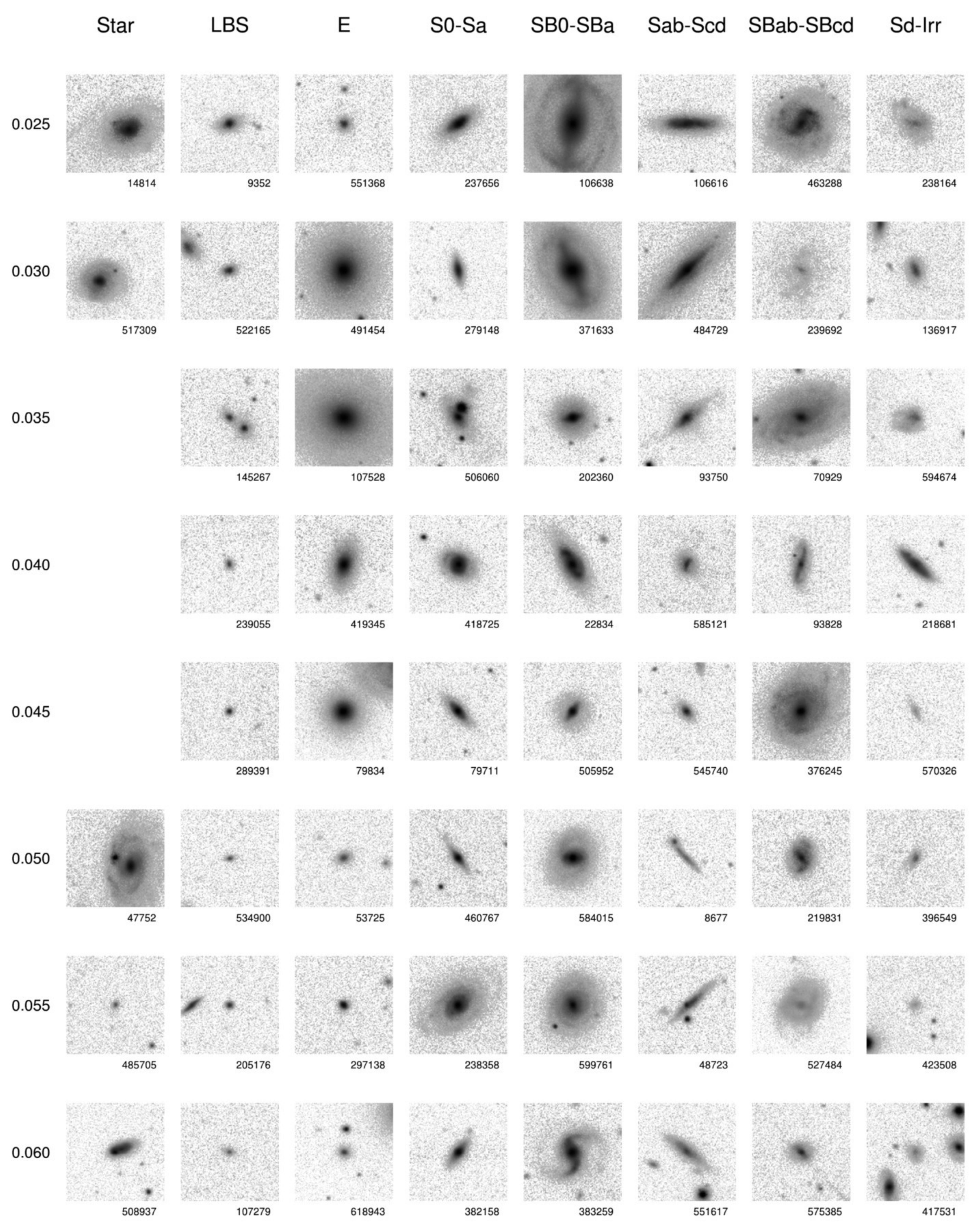}

\caption{\label{fig:zmorpheye}Example postage-stamp cutouts for each morphological
class, arranged according to redshift. Below each postage-stamp is
the GAMA ID of the galaxy, for reference. The images shown here are
created from arctan-scaled composite three-colour images (RGB taken
from $Hig$, respectively), with the colours desaturated and inverted
to create a greyscale black-on-white image. Blank spaces show regions
where no objects of that class exist.}
\end{figure*}

Example greyscale postage-stamp images for the various visual morphological
classes are shown in Figure \ref{fig:zmorpheye}, arranged according
to local flow corrected redshift. The Star and LBS classes are included
here for reference. A comparison between our own morphological classifications
and those of Galaxy Zoo can be found in Appendix \ref{app:galaxyzoocomparison},
and further $3$-colour postage-stamp examples for each morphological
class arranged into a colour-Sérsic index plane may be found in Appendix
\ref{app:morphologies}.

\section{Morphological Trends}

\label{sec:trends}

\subsection{Trends With Global Properties}

\label{sub:globaltrends}
\begin{figure*}
\includegraphics[width=1\textwidth]{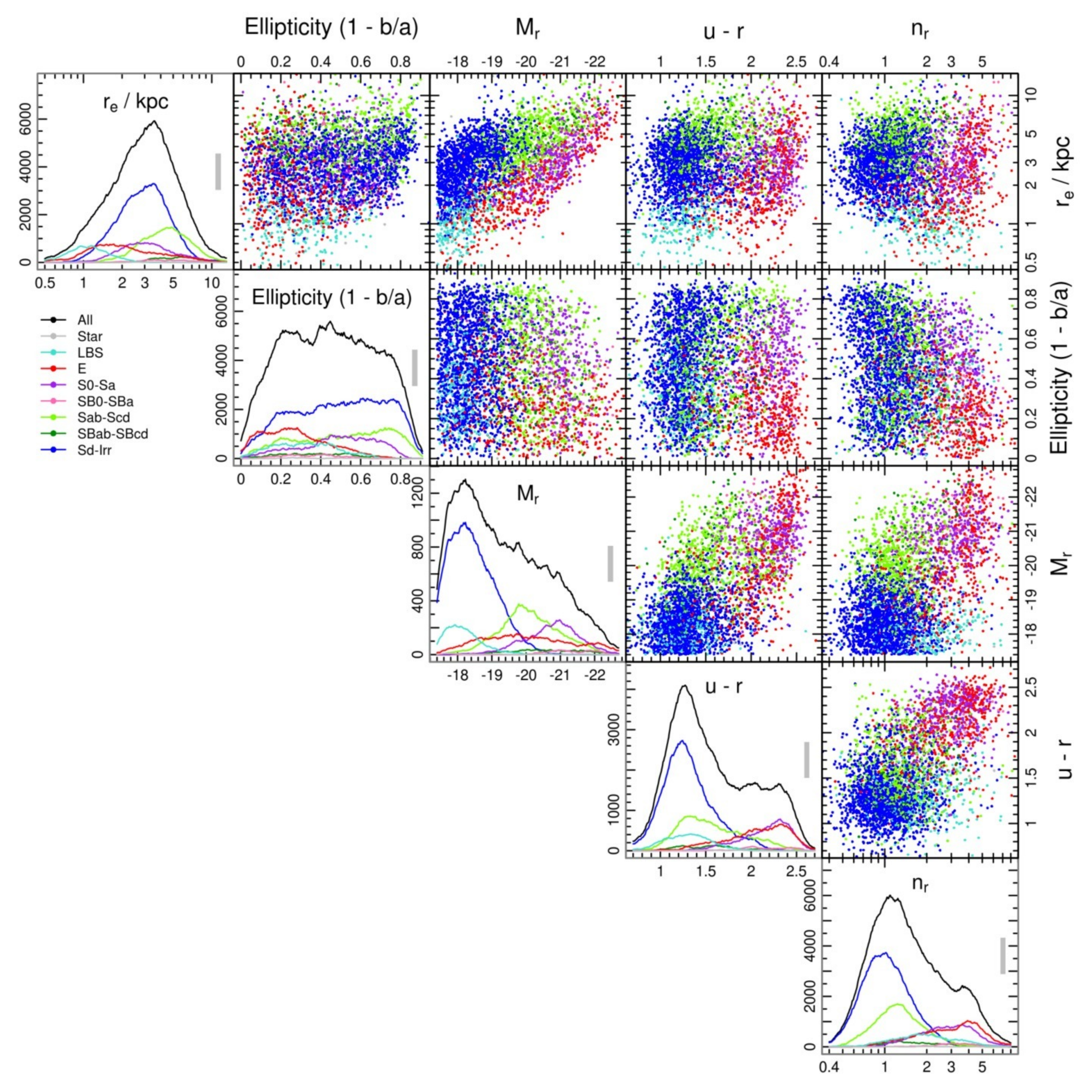}

\caption{\label{fig:allvsall}Correlation matrix showing five global parameters,
namely (from left to right): $r$ band half-light radius (in kpc),
ellipticity, absolute $r$ band Sérsic magnitude, $(u-r)$ rest frame
colour from SED fitting and $r$ band Sérsic index. The associated
1D density plots have been constructed using rectangular bandwidth
standard deviations of $0.15$, $0.09$, $0.55$, $0.2$ and $0.13$,
respectively, as indicated by the width of the grey rectangles inset
into each density sub-plot. The density plots integrate to the total
number of objects in each population. Data points are coloured according
to their visual morphological classification, as indicated. Distinct
groupings of similar colour data points (i.e., same morphology) can
be seen, particularly in the case of absolute magnitude against half-light
radius where the red-sequence for elliptical galaxies and the blue
cloud for Sd-Irr type galaxies is clearly visible.}
\end{figure*}

In Figure \ref{fig:allvsall} we show five
global galaxy measurements against one another, coloured according
to their morphological classification. The five measurements shown
are $r$ band measured half-light radius (kpc), ellipticity as measured
in the $r$ band, absolute $r$ band Sérsic magnitude (truncated at
$10r_{e}$), rest-frame $(u-r)$ colour from the best-fitting SED
and $r$ band Sérsic index. Sérsic measurements come about from a
single-Sérsic fit to the $r$ band data \citep{Kelvin2012}. Absolute
Sérsic magnitudes are calculated in the standard sense, using Equation
\ref{eq:abmag}.

It can be seen that some projections of the data more easily allow
distinct morphological groupings to be brought out than others. Absolute
magnitude against half-light radius shows a red-sequence of elliptical-type
galaxies progressing from the bright extended end diagonally downwards
towards the compact faint region of the figure, slightly exhibiting
the curvature that is known to become more apparent at magnitudes
fainter than that sampled here (e.g., \citealp{Binggeli1984,Graham2008b,Forbes2008,Misgeld2011}).
Note the elliptical galaxy extension of this curved relation in L-r$_{e}$
space directly into the LBS regime. Clear bimodalities in the data
can be seen in the planes of absolute magnitude vs $(u-r)$ colour,
absolute magnitude vs Sérsic index, and $(u-r)$ colour vs Sérsic
index. As has been shown in, e.g., \citet{Baldry2004,Driver2006,Conselice2006b},
these bimodal distributions are well fitted by a double-Gaussian profile.

Spheroid-dominated bulge+disk systems (S0-Sa/SB0-SBa) all occupy the
same parameter space as the single-component elliptical galaxies,
lying on top of the red-sequence. These results are in good agreement
with the conclusions of \citet{Drory2007}, who find that spiral galaxies
harbouring classical bulges lie consistently on the red-sequence.
However, we do not assert that all S0-Sa galaxies harbour a classical
bulge, as has previously been shown for the S0 type galaxy NGC $2787$
\citep{Erwin2003}. Although speculation remains as to how $z=0$
classical bulges came into existence, one hypothesis \citep{Driver2013}
is that they may have formed from the compact elliptical galaxies
at $z\sim2\pm0.5$. The evolutionary path of these high-redshift compact
galaxies may have diverged from today's classical bulges, having grown
a disk through gas accretion (e.g., \citealp{Navarro1991,Steinmetz2002,Graham2013})
while today's elliptical galaxies puffed-up via progressive minor
accretion events (see \citealp{Driver2013}, and references therein),
but see \citet{Carollo2013}. In contrast to classical bulges, pseudo-bulges
are believed to form via secular evolutionary processes present within
the disk \citep{Debattista2006,Gadotti2009,Saha2012}. In brief: if
left in isolation for a sufficient length of time (i.e., without any
major merging events), a dynamically cold rotating disk system will
form a barred structure. In practice, external gravitational triggers
(flybys, rather than mergers) are additionally responsible for inducing
the formation of bars. The bar acts as a very efficient means by which
stellar mass and gas in the disk may be funnelled into the core of
the galaxy, initiating a new phase of star formation in the central
region. A young, blue sub-structure exhibiting a large component of
angular velocity and a flattened 2D-like structure with a low central
concentration (Sérsic $n\lesssim2$) will form. This new structure
is commonly referred to as a pseudo-bulge. We note however that it
is possible to form low Sérsic index ($n<2$) bulges via other non-secular
processes (see a full review in \citealp{Graham2013}). Unlike classical
bulges, \citeauthor{Drory2007} find that galaxies with pseudo-bulges
typically lie in the blue cloud \citep{Drory2007}. We find very few
multi-component systems overlapping with the main body of the blue
cloud, and conclude that structural decomposition is required in order
to comment further on a) which of these galaxies may contain a pseudo-bulge
and b) where these galaxies lie in relation to the blue cloud.

\subsection{Trends With Redshift}

\label{sub:redshifttrends}
\begin{figure}
\includegraphics[width=1\columnwidth]{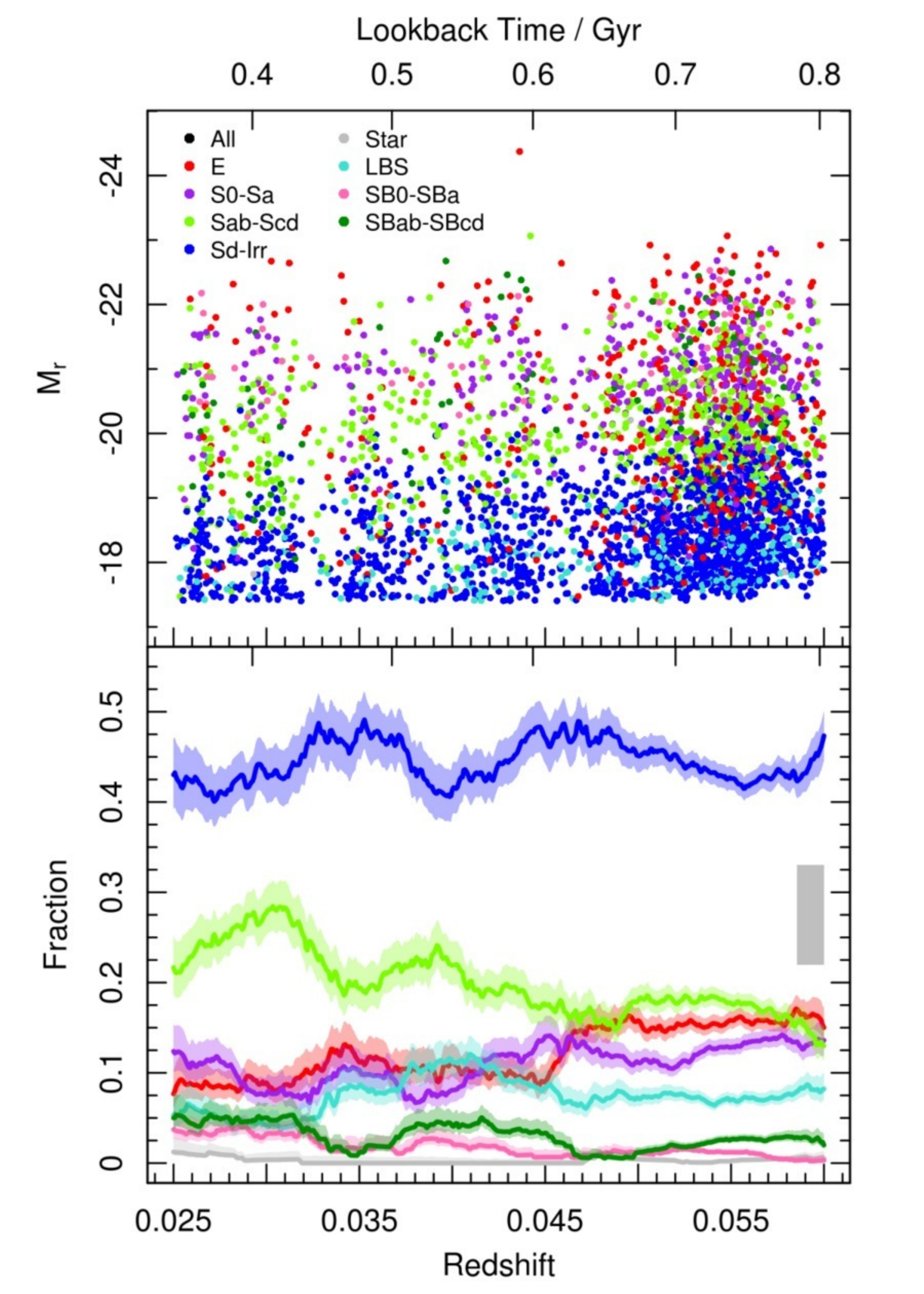}

\caption{\label{fig:zmorph}Morphology against redshift. Data points are coloured
according to their morphology, as indicated. (top) Absolute $r$ band
Sérsic magnitude as a function of redshift. (bottom) The representative
fractions of the total number of galaxies for each morphology, as
a function of redshift. Shaded regions indicate $\pm1\sigma$ binomial
confidence intervals \citep{Cameron2011}. These fractions have been
constructed using a rectangular kernel with a bandwidth standard deviation
of $0.005$, as indicated by the width of the grey rectangle inset
into the lower plot.}
\end{figure}

One would not expect to see large evolutionary
variations in morphological fraction over a narrow redshift range
such as that used in the creation of our volume limited sample. Figure
\ref{fig:zmorph} shows the data as a function of redshift, with data
points coloured according to their morphology, as indicated. These
data are shown relative to their absolute $r$ band Sérsic magnitudes
(top) and number fractions (bottom). Shaded regions indicate $\pm1\sigma$
binomial confidence intervals \citep{Cameron2011}. One can clearly
see the large-scale structure with redshift appearing as over-dense
strips in the scatter plot. The two distinctive peaks in the Sd-Irr
type galaxy population at redshifts of low density ($z\sim0.035$
and $z\sim0.047$) reflect the aforementioned morphology-density relation
of \citet{Dressler1980}.

Elliptical galaxies (red) exhibit a minor fractional evolution over
this redshift range, with a higher proportion of elliptical galaxies
at the high redshift end of our sample relative to our lowest redshift
bin. Moving from high to low redshift, as the elliptical fraction
drops off it is replaced by Sab-Scd (and, to a lesser extent, SBab-SBcd)
type galaxies. However, these trends appear to be minor, and confirm
that these data do not show large evolutionary variations in morphological
fraction with redshift. Note that we did observe distinct fractional
evolutionary trends at redshifts below $z=0.025$, i.e., our lower
limit, however, we give low credence to these results owing to the
very low numbers of galaxies in our volume at $z<0.025$ ($\sim300$
galaxies in the redshift range $0.013<z<0.025$).

\section{Luminosity Functions}

\label{sec:lumfuncs}The luminosity function (LF) describes the number
density of galaxies in any given luminosity (or magnitude) bin across
a wide range of luminosities. Measurement of the LF allows for constraints
to be placed on galaxy formation and evolution models, and as such
is valuable and informative.

\subsection{The Schechter Luminosity Function}

The \citet{Schechter1976} Luminosity Function is an analytical representation
of the luminosity function, describing the number of galaxies per
unit volume in the luminosity interval $L$ to $L+\mathrm{d}L$, where
$dL$ is some linear luminosity interval. The number density, $\phi\left(L\right)\mathrm{d}L=\mathrm{d}n$,
is given by
\begin{equation}
\phi\left(L\right)\mathrm{d}L=\phi^{*}\left(\frac{L}{L^{*}}\right)^{\alpha}\exp\left(-\frac{L}{L^{*}}\right)\frac{1}{L^{*}}\mathrm{d}L\label{eq:schechter}
\end{equation}
where $\phi^{*}$ is the normalisation constant, $L^{*}$ is the characteristic
luminosity describing the position of the `knee' in the luminosity
function and $\alpha$ gives the slope of the luminosity function
at the faint end (where $L\ll L^{*}$). Note that $\phi^{*}$, $L^{*}$
and $\alpha$ are to be determined by minimising a fit to the data.
The impact of the Schechter function is to truncate the bright-end
power law distribution of galaxies, vastly reducing number counts
at luminosities brighter than $L^{*}$. 

It is usually more convenient when considering luminosities to re-write
the Schechter Function in terms of magnitude, as given by

\begin{multline}
\Phi\left(M\right)\mathrm{d}M=0.4\ln10\cdot\phi^{*}10^{-0.4(M-M^{*})(\alpha+1)}\\
\times\exp\left(-10^{-0.4(M-M^{*})}\right)\mathrm{d}M\label{eq:schechter-mag}
\end{multline}

where $M$ and $M^{*}$ are the magnitude and the characteristic magnitude
corresponding to $L^{*}$. The parameters $\alpha$ and $\phi^{*}$
now correspond to the slope and normalisation constant in magnitude
space. Equation \ref{eq:schechter-mag} is the form of the Schechter
function we shall assume throughout the remainder of this paper.

\subsection{Measuring the Luminosity Function}

\label{sub:morphologyluminosityfunction}The total galaxy luminosity
function (GLF) is dominated by large numbers of very faint galaxies,
while the space density of brighter galaxies drops off sharply beyond
some given luminosity ($L^{*}$). Despite their numerical dominance
however, low luminosity systems tend to contribute a relatively small
fraction to the total luminosity budget of any given volume \citep{Driver1999}.
The GLF is a combination of its constituent morphological-type luminosity
functions (MLFs), with each morphological type contributing variable
number densities, dependent upon magnitude.

\begin{figure*}
\includegraphics[width=1\textwidth]{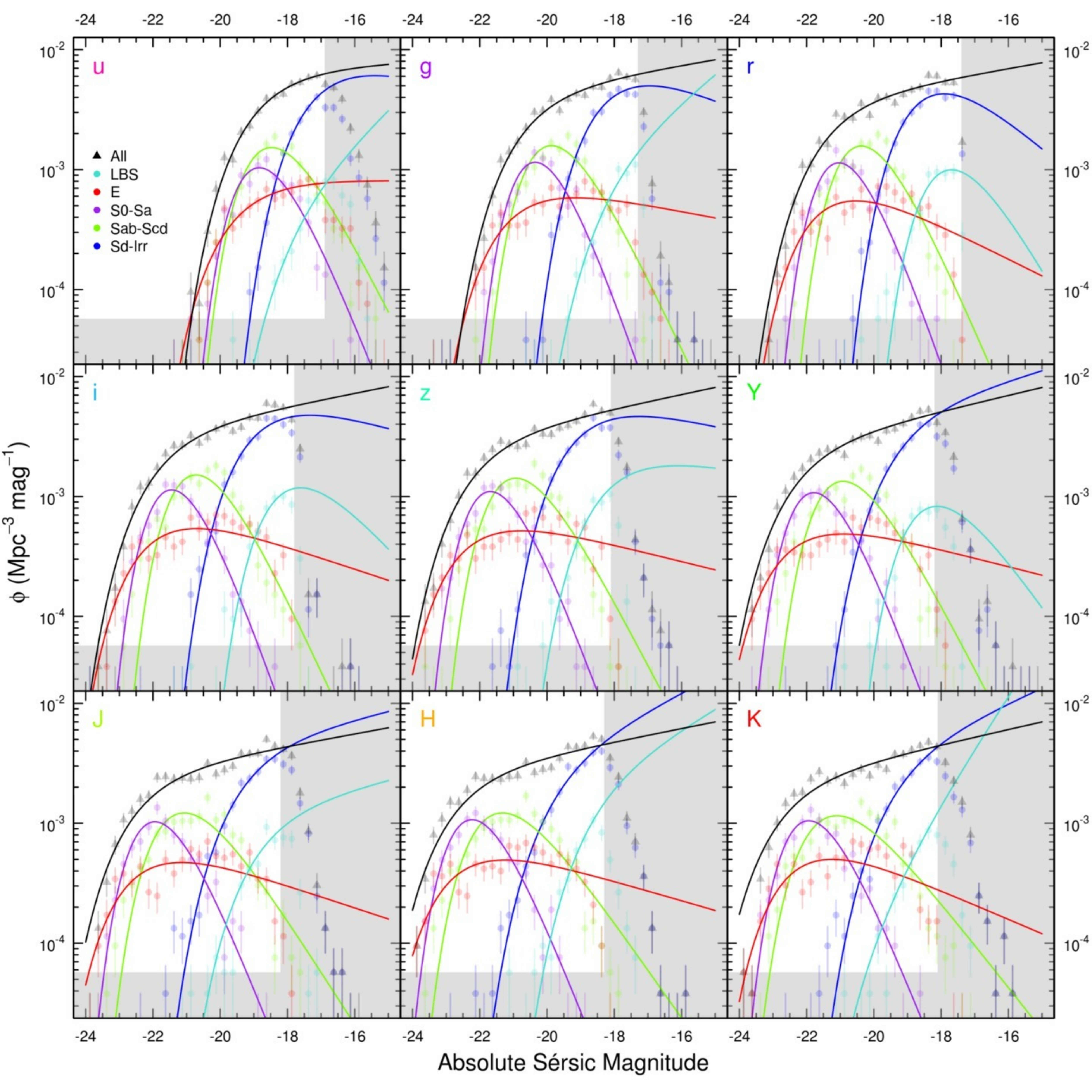}

\caption{\label{fig:lumfuncmultisingle}Morphological-type Luminosity Functions
across all nine bands for the various morphological types (coloured
points and lines, as indicated) and total populations (black points
and lines). Each population has been fit with a single Schechter function.
Prior to fitting, the data are split into bins of $0.25$ mag, with
the error on the measurement per bin taken as Poissonian ($\sqrt{n}$)
in nature. Shaded grey areas indicate those regions where data has
not been used in the fits. Variable faint-end magnitude limits are
given in Table \ref{tab:faintlimits}.}
\end{figure*}

Figure \ref{fig:lumfuncmultisingle} shows the GLF and constituent
MLFs across nine GAMA wavelengths ($ugrizYJHK$) for our volume-limited
GAMAnear sample of $3,727$ galaxies. The barred populations have
been merged into their sibling classes owing to low number statistics
for those two populations. Each population (total and morphological
type) is binned into absolute Sérsic magnitude bins of $0.25$ mag
and fit with a single Schechter function%
\footnote{Schechter functions are fit to the available data within our magnitude
ranges using the \noun{nlminb} routine in R; a quasi-Newton algorithm
based on the PORT routines that optimise fitting in a similar sense
to the Limited-memory Broyden-Fletcher-Goldfarb-Shanno algorithm (LM-BFGS),
with an extension to handle simple box constraints on input variables
(L-BFGS-B). The PORT documentation is available at http://netlib.bell-labs.com/cm/cs/cstr/153.pdf.%
}. Errors on each bin are assumed to be Poissonian. Shaded grey areas
indicate the limits of the fit beyond which data were not used to
constrain the Schechter function model. These limits are a minimum
number count of $n\le3$ and an absolute Sérsic magnitude faint-end
cutoff as given in Table \ref{tab:faintlimits}.

The knee in the total Schechter function progresses smoothly towards
brighter AB magnitudes as one moves from $u$ to $K$, as expected.
We find the knee to be generally well fitted with a single Schechter
function until $\sim z$ band. At longer wavelengths, the GLF appears
to require a secondary component to aid in fully reproducing the downturn
at the bright end and the secondary upturn at the faint end. 

These data provide one of the first measurements of the MLF using
Sérsic photometry, and provide a key insight into the nature of the
underlying galaxy populations. Considering the morphology sub-populations
alone, the faint end appears to be heavily dominated by Sd-Irr type
galaxies, in addition to a significant LBS fraction. Intermediate
magnitudes typically contain both the S0-Sa and Sab-Scd type systems.
Elliptical galaxies dominate at the brightest magnitudes, however,
below their $L^{*}$ knee the number of E-type galaxies remains relatively
constant across all wavelengths. The Sd-Irr and LBS populations appear
to show the largest variation in their MLFs with respect to wavelength,
with the faint end slopes varying strongly from $u$ to $K$ as the
relative depth of the data in those bands becomes shallower. Owing
to our sample selection constraints and the relatively high quality
of the $r$ band data, one would expect the $M^{*}$ and $\alpha$
parameters for Sd-Irr and LBS type galaxies in the $r$ band to be
the most robust in relation to other passbands, which is perhaps evidenced
by the suggestion of a downturn in the Sd-Irr type galaxy population
at the faintest ($M_{r}>-18.5$ mag) magnitudes. In contrast, the
E, S0-Sa and Sab-Scd populations reproduce consistent luminosity function
curves at each wavelength, albeit offset in magnitude.

Single Schechter fit parameters are shown for all populations in Tables
\ref{tab:lumfuncmultisingle} (All); \ref{tab:lumfuncmultiLBS} (LBS);
\ref{tab:lumfuncmultiE} (E); \ref{tab:lumfuncmultiS0a} (S0-Sa);
\ref{tab:lumfuncmultiSbc} (Sab-Scd) and \ref{tab:lumfuncmultiSd}
(Sd-Irr). In addition to the Schechter fit parameters, we also calculate
the luminosity density for each population at each wavelength. The
luminosity density, $j$, is the integral under the Schechter function
curve and is given by 
\[
j=\intop_{0}^{\infty}L\phi(L)\mathrm{d}L=\phi^{*}L^{*}\Gamma(\alpha+2)
\]
as in \citet{Liske2003b}. Note that the luminosity densities are
those calculated from an extrapolation across all luminosities. Also
note that the quoted errors on $j$ are likely to be a lower bound
owing to the correlation of errors in $L^{*}$, $\alpha$ and $\phi^{*}$.

Alongside the characteristic knee in the Schechter luminosity function,
$L^{*}$ (or $M^{*}$), the remaining fitted parameters are the slope
of the faint end of the LF, $\alpha$, and the normalisation $\phi^{*}$.
While the error on the latter may be estimated via some simplistic
method such as jackknife resampling%
\footnote{A statistical resampling method designed to estimate sample bias and
variance by systematically recomputing our Schechter fit parameters
on numerous subsets of our data.%
} of the data set, the well-known covariance between $\alpha$ and
$M^{*}$ would result in their jackknife errors being systematically
underestimated. An alternative approach is to produce error ellipses
which map out the $\chi^{2}$ parameter space around the best fit
values. This technique involves re-fitting the data set fitting for
$\phi^{*}$ alone while adopting a fixed pair of input $\alpha$ and
$M^{*}$ parameters as defined by a regularly spaced grid about the
best fit values. Assuming Gaussian errors, the resultant $\chi^{2}$
surface then allows for $1$, $2$ and $3\sigma$ errors to be determined
as the contours which lie at $\Delta\chi^{2}=2.30$, $6.17$ and $11.8$,
respectively. 

\begin{figure*}
\includegraphics[width=1\textwidth]{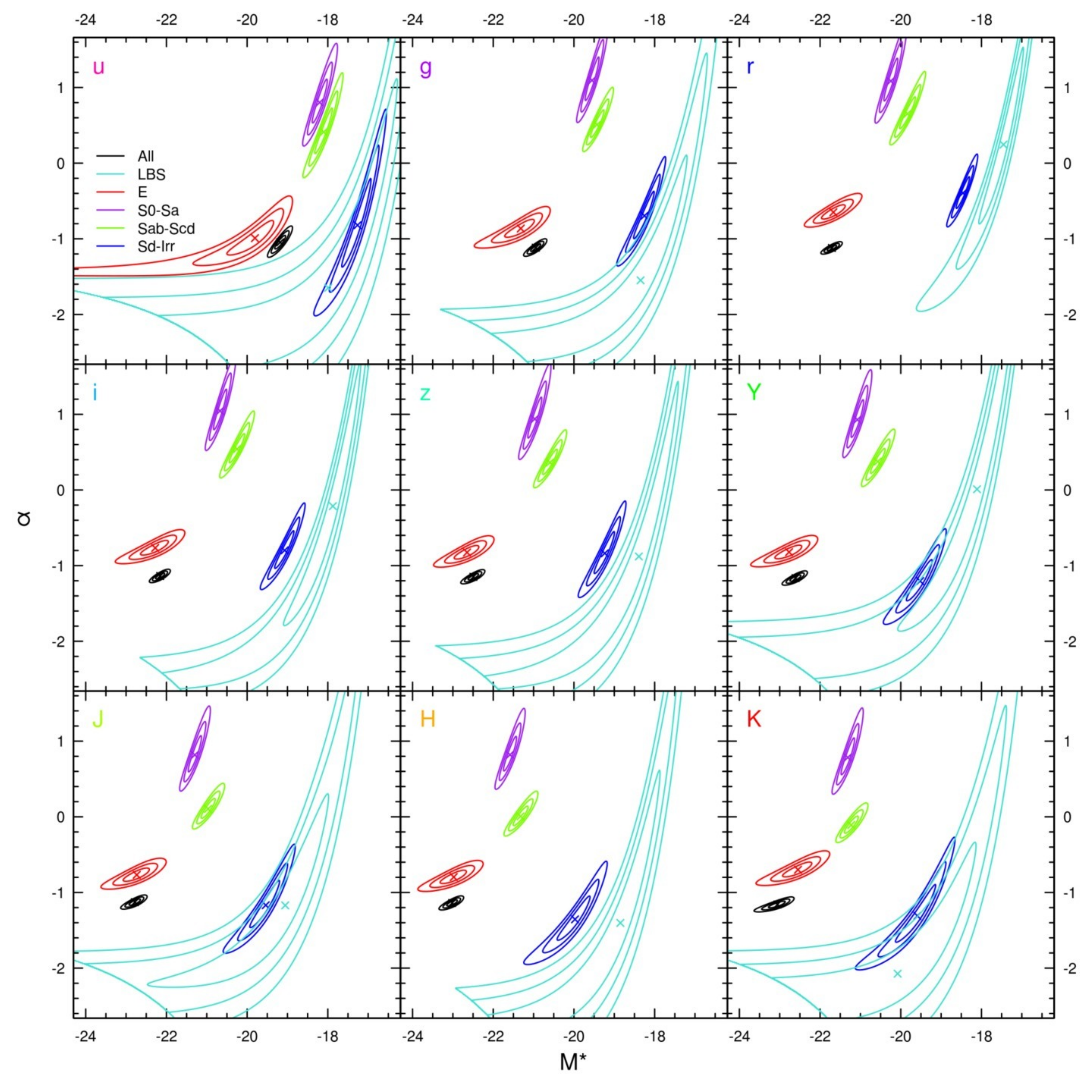}

\caption{\label{fig:lumfuncmultisingleellipses}Error ellipses for each Schechter
function fit shown in Figure \ref{fig:lumfuncmultisingle}. These
ellipses are generated by constructing a regularly spaced grid of
input $M^{*}$ and $\alpha$ values in steps of $0.01$ each and fitting
for the normalisation constant $\phi^{*}$ in the Schechter function,
producing a $\chi^{2}$ map about the coordinates of the best fit.
Successive contours represent the $1\sigma$, $2\sigma$ and $3\sigma$
error boundaries ($\Delta\chi^{2}=2.30$, $6.17$ and $11.8$ respectively).
Note the significant diagonal elongation between these parameters,
particularly for the Sd class population. This highlights the covariant
relationship between $M^{*}$ and $\alpha$.}
\end{figure*}

We thus derive error ellipses for all nine photometric bands of our
dataset, and show the results in Figure \ref{fig:lumfuncmultisingleellipses}.
Error ellipses for the total GLF and the constituent MLFs are shown,
as indicated. Successive contours represent the $1$, $2$ and $3\sigma$
errors on each parameter. As can be seen, the strong covariance between
these two parameters has a strong impact on each error ellipse. The
LBS population shows the largest errors, which should not be surprising
considering our sample selection limits and the typically faint magnitude
of these systems. The visible truncation of the LBS error ellipses
toward the bright end of each figure (with the exception of the $r$
band) is as expected, owing to a lack of LBS type systems at brighter
magnitudes. The brightest LBS in our sample has a Sérsic $r$ band
magnitude of $M_{r}=-20.82$ mag. Only the $r$ band data allow any
meaningful constraints to be placed on Schechter fit parameters to
the LBS population (and to some extent, the Sd population also), particularly
in constraining $M^{*}$.

Of the standard Hubble types, the faint end slope of the Sd-Irr class
in the $u$ band, $\alpha_{u}=-0.82_{-0.55}^{+0.61}$, is particularly
poorly constrained owing to the poor quality and relatively shallow
depth of the $u$ band in conjunction with the completeness issues
for the Sd-Irr population. Conversely, while $\alpha$ is typically
well constrained for the elliptical populations, the value of the
knee in the Schechter function is not. In the $K$ band for example,
the turnover is found at $M_{K}^{*}=-22.55_{-0.38}^{+0.34}$ mag;
a relatively large uncertainty. Also note the relative consistency
between recovered $\alpha$ values for all populations over all wavelengths,
excepting the $u$ band and LBS populations as discussed above.

\renewcommand{\arraystretch}{1.5}
\setlength{\tabcolsep}{1pt}

\begin{table}
\begin{centering}
\begin{tabular}{cccccc}
\hline 
Band & $M^{*}$ & $\alpha$ & $\phi^{*}/10^{-3}$ & $\chi^{2}/\nu$ & $j/10^{7}$\tabularnewline
 & (mag) &  & ($\mathrm{mag}^{-1}\mathrm{Mpc}^{-3}$) &  & ($\mathrm{L}_{\odot}\mathrm{Mpc}^{-3}$)\tabularnewline
\hline 
$u$ & $-19.18_{-0.13}^{+0.13}$ & $-1.05_{-0.08}^{+0.09}$ & $6.99\pm1.60$ & $1.12$ & $12.05_{-0.86}^{+0.86}$ \tabularnewline
$g$ & $-20.95_{-0.12}^{+0.12}$ & $-1.12_{-0.04}^{+0.05}$ & $4.71\pm0.37$ & $1.92$ & $14.00_{-0.51}^{+0.51}$ \tabularnewline
$r$ & $-21.71_{-0.11}^{+0.11}$ & $-1.12_{-0.03}^{+0.03}$ & $4.00\pm0.21$ & $2.79$ & $16.02_{-0.81}^{+0.81}$ \tabularnewline
$i$ & $-22.15_{-0.12}^{+0.10}$ & $-1.14_{-0.03}^{+0.04}$ & $3.61\pm0.52$ & $2.46$ & $19.29_{-1.39}^{+1.39}$ \tabularnewline
$z$ & $-22.49_{-0.13}^{+0.12}$ & $-1.15_{-0.03}^{+0.04}$ & $3.17\pm0.22$ & $2.32$ & $22.89_{-1.30}^{+1.30}$ \tabularnewline
$Y$ & $-22.61_{-0.13}^{+0.13}$ & $-1.16_{-0.04}^{+0.03}$ & $2.77\pm0.37$ & $2.25$ & $22.16_{-1.16}^{+1.16}$ \tabularnewline
$J$ & $-22.78_{-0.14}^{+0.13}$ & $-1.13_{-0.03}^{+0.04}$ & $2.72\pm0.27$ & $3.18$ & $25.90_{-1.64}^{+1.64}$ \tabularnewline
$H$ & $-23.02_{-0.13}^{+0.13}$ & $-1.14_{-0.03}^{+0.04}$ & $2.73\pm0.63$ & $4.32$ & $37.08_{-7.94}^{+7.94}$ \tabularnewline
$K$ & $-23.06_{-0.21}^{+0.19}$ & $-1.16_{-0.04}^{+0.04}$ & $2.33\pm0.39$ & $3.11$ & $52.07_{-5.63}^{+5.63}$ \tabularnewline

\hline 
\end{tabular}
\par\end{centering}

\caption{\label{tab:lumfuncmultisingle}Single Schechter luminosity function
fit parameters for the total GLF as shown in Figure \ref{fig:lumfuncmultisingle}.
From left to right, columns are: GAMA passband; the knee in the Schechter
function ($M^{*}$); the slope of the faint end of the Schechter function
($\alpha$); the normalisation constant of the Schechter function
($\phi^{*}$); the $\chi^{2}$ goodness of fit parameter ($\chi^{2}/\nu$),
and; the luminosity density ($j$). Errors on $M^{*}$ and $\alpha$
are taken from the $1\sigma$ error ellipses shown in Figure \ref{fig:lumfuncmultisingleellipses}.
All other errors are estimated using the relation $\sigma^{2}=\frac{N-1}{N}\sum_{i=1}^{N}\left(x_{j}-x\right)^{2}$,
where $x$ is the best fit parameter, $x_{j}$ is the best fit parameter
as given from a jackknife resampled variant of the data set and $N$
is the number of jackknife volumes. We adopt $N=10$.}
\end{table}

\renewcommand{\arraystretch}{1}

\renewcommand{\arraystretch}{1.5}

\begin{table}
\begin{centering}
\begin{tabular}{cccccc}
\hline 
Band & $M^{*}$ & $\alpha$ & $\phi^{*}/10^{-3}$ & $\chi^{2}/\nu$ & $j/10^{7}$\tabularnewline
 & (mag) &  & ($\mathrm{mag}^{-1}\mathrm{Mpc}^{-3}$) &  & ($\mathrm{L}_{\odot}\mathrm{Mpc}^{-3}$)\tabularnewline
\hline 
$u$ & $-17.99_{-4.22}^{+1.71}$ & $-1.65_{-1.01}^{+2.76}$ & $0.60\pm1.98$ & $0.47$ & $0.85_{-0.85}^{+4.60}$ \tabularnewline
$g$ & $-18.35_{-3.70}^{+1.15}$ & $-1.55_{-1.07}^{+1.65}$ & $1.29\pm2.03$ & $0.93$ & $0.63_{-0.63}^{+2.27}$ \tabularnewline
$r$ & $-17.45_{-0.61}^{+0.49}$ & $0.25_{-1.05}^{+1.28}$ & $2.87\pm1.21$ & $1.10$ & $0.24_{-0.03}^{+0.03}$ \tabularnewline
$i$ & $-17.88_{-1.21}^{+0.69}$ & $-0.21_{-1.58}^{+1.87}$ & $3.40\pm2.20$ & $1.32$ & $0.30_{-0.12}^{+0.12}$ \tabularnewline
$z$ & $-18.39_{-3.84}^{+0.97}$ & $-0.88_{-1.75}^{+2.31}$ & $2.85\pm2.97$ & $1.10$ & $0.40_{-0.40}^{+0.93}$ \tabularnewline
$Y$ & $-18.11_{-1.96}^{+0.75}$ & $0.01_{-1.87}^{+1.65}$ & $2.45\pm2.62$ & $1.47$ & $0.28_{-0.12}^{+0.12}$ \tabularnewline
$J$ & $-19.06_{-3.41}^{+1.07}$ & $-1.17_{-1.08}^{+1.47}$ & $1.33\pm2.13$ & $1.72$ & $0.43_{-0.43}^{+1.15}$ \tabularnewline
$H$ & $-18.85_{-3.29}^{+0.98}$ & $-1.40_{-1.26}^{+2.01}$ & $2.38\pm3.90$ & $0.71$ & $0.94_{-0.94}^{+12.61}$ \tabularnewline
$K$ & $-20.08_{-3.14}^{+1.94}$ & $-2.07_{-0.59}^{+1.73}$ & $0.22\pm1.31$ & $0.21$ & $-_{}^{}$ \tabularnewline

\hline 
\end{tabular}
\par\end{centering}

\caption{\label{tab:lumfuncmultiLBS}As Table \ref{tab:lumfuncmultisingle}
but for little blue spheroids (LBS).}
\end{table}

\renewcommand{\arraystretch}{1}

\renewcommand{\arraystretch}{1.5}

\begin{table}
\begin{centering}
\begin{tabular}{cccccc}
\hline 
Band & $M^{*}$ & $\alpha$ & $\phi^{*}/10^{-3}$ & $\chi^{2}/\nu$ & $j/10^{7}$\tabularnewline
 & (mag) &  & ($\mathrm{mag}^{-1}\mathrm{Mpc}^{-3}$) &  & ($\mathrm{L}_{\odot}\mathrm{Mpc}^{-3}$)\tabularnewline
\hline 
$u$ & $-19.81_{-0.69}^{+0.46}$ & $-0.99_{-0.22}^{+0.23}$ & $0.92\pm0.41$ & $0.67$ & $2.73_{-1.10}^{+1.10}$ \tabularnewline
$g$ & $-21.32_{-0.39}^{+0.33}$ & $-0.86_{-0.12}^{+0.11}$ & $0.95\pm0.21$ & $0.96$ & $3.44_{-0.29}^{+0.29}$ \tabularnewline
$r$ & $-21.67_{-0.29}^{+0.29}$ & $-0.65_{-0.08}^{+0.10}$ & $1.22\pm0.19$ & $2.41$ & $3.88_{-0.37}^{+0.37}$ \tabularnewline
$i$ & $-22.28_{-0.36}^{+0.32}$ & $-0.77_{-0.09}^{+0.10}$ & $1.04\pm0.15$ & $1.58$ & $5.13_{-0.70}^{+0.70}$ \tabularnewline
$z$ & $-22.65_{-0.31}^{+0.29}$ & $-0.83_{-0.08}^{+0.09}$ & $0.91\pm0.14$ & $1.60$ & $6.29_{-0.43}^{+0.43}$ \tabularnewline
$Y$ & $-22.77_{-0.35}^{+0.31}$ & $-0.82_{-0.08}^{+0.09}$ & $0.86\pm0.17$ & $1.17$ & $6.53_{-0.50}^{+0.50}$ \tabularnewline
$J$ & $-22.73_{-0.34}^{+0.30}$ & $-0.77_{-0.08}^{+0.09}$ & $0.91\pm0.22$ & $2.27$ & $6.89_{-0.75}^{+0.75}$ \tabularnewline
$H$ & $-22.98_{-0.33}^{+0.31}$ & $-0.80_{-0.08}^{+0.09}$ & $0.91\pm0.15$ & $1.91$ & $9.91_{-0.98}^{+0.98}$ \tabularnewline
$K$ & $-22.55_{-0.38}^{+0.34}$ & $-0.70_{-0.09}^{+0.10}$ & $1.05\pm0.21$ & $3.07$ & $11.76_{-1.96}^{+1.96}$ \tabularnewline

\hline 
\end{tabular}
\par\end{centering}

\caption{\label{tab:lumfuncmultiE}As Table \ref{tab:lumfuncmultisingle} but
for elliptical galaxies.}
\end{table}

\renewcommand{\arraystretch}{1}

\renewcommand{\arraystretch}{1.5}

\begin{table}
\begin{centering}
\begin{tabular}{cccccc}
\hline 
Band & $M^{*}$ & $\alpha$ & $\phi^{*}/10^{-3}$ & $\chi^{2}/\nu$ & $j/10^{7}$\tabularnewline
 & (mag) &  & ($\mathrm{mag}^{-1}\mathrm{Mpc}^{-3}$) &  & ($\mathrm{L}_{\odot}\mathrm{Mpc}^{-3}$)\tabularnewline
\hline 
$u$ & $-18.20_{-0.18}^{+0.18}$ & $0.80_{-0.26}^{+0.31}$ & $2.37\pm0.24$ & $1.53$ & $2.71_{-0.22}^{+0.22}$ \tabularnewline
$g$ & $-19.55_{-0.17}^{+0.17}$ & $1.09_{-0.26}^{+0.31}$ & $2.17\pm0.19$ & $1.45$ & $3.61_{-0.28}^{+0.28}$ \tabularnewline
$r$ & $-20.25_{-0.16}^{+0.17}$ & $1.08_{-0.25}^{+0.30}$ & $2.16\pm0.29$ & $1.60$ & $4.50_{-0.34}^{+0.34}$ \tabularnewline
$i$ & $-20.67_{-0.16}^{+0.16}$ & $1.04_{-0.24}^{+0.29}$ & $2.22\pm0.25$ & $1.18$ & $5.72_{-0.49}^{+0.49}$ \tabularnewline
$z$ & $-20.97_{-0.17}^{+0.17}$ & $0.94_{-0.25}^{+0.28}$ & $2.31\pm0.18$ & $1.23$ & $6.99_{-0.51}^{+0.51}$ \tabularnewline
$Y$ & $-21.08_{-0.15}^{+0.15}$ & $0.93_{-0.23}^{+0.26}$ & $2.27\pm0.32$ & $0.71$ & $7.39_{-0.62}^{+0.62}$ \tabularnewline
$J$ & $-21.30_{-0.16}^{+0.16}$ & $0.82_{-0.22}^{+0.25}$ & $2.33\pm0.37$ & $1.25$ & $8.82_{-0.80}^{+0.80}$ \tabularnewline
$H$ & $-21.58_{-0.16}^{+0.15}$ & $0.82_{-0.21}^{+0.24}$ & $2.43\pm0.32$ & $0.87$ & $13.63_{-1.36}^{+1.36}$ \tabularnewline
$K$ & $-21.31_{-0.17}^{+0.16}$ & $0.78_{-0.22}^{+0.26}$ & $2.43\pm0.17$ & $1.03$ & $16.02_{-1.19}^{+1.19}$ \tabularnewline

\hline 
\end{tabular}
\par\end{centering}

\caption{\label{tab:lumfuncmultiS0a}As Table \ref{tab:lumfuncmultisingle}
but for S(B)0-S(B)a galaxies.}
\end{table}

\renewcommand{\arraystretch}{1}

\renewcommand{\arraystretch}{1.5}

\begin{table}
\begin{centering}
\begin{tabular}{cccccc}
\hline 
Band & $M^{*}$ & $\alpha$ & $\phi^{*}/10^{-3}$ & $\chi^{2}/\nu$ & $j/10^{7}$\tabularnewline
 & (mag) &  & ($\mathrm{mag}^{-1}\mathrm{Mpc}^{-3}$) &  & ($\mathrm{L}_{\odot}\mathrm{Mpc}^{-3}$)\tabularnewline
\hline 
$u$ & $-18.10_{-0.22}^{+0.21}$ & $0.41_{-0.28}^{+0.31}$ & $4.20\pm0.27$ & $2.03$ & $3.27_{-0.28}^{+0.28}$ \tabularnewline
$g$ & $-19.43_{-0.15}^{+0.17}$ & $0.51_{-0.17}^{+0.19}$ & $4.19\pm0.16$ & $3.20$ & $3.80_{-0.37}^{+0.37}$ \tabularnewline
$r$ & $-19.83_{-0.19}^{+0.19}$ & $0.65_{-0.19}^{+0.22}$ & $3.90\pm0.30$ & $4.54$ & $3.81_{-0.47}^{+0.47}$ \tabularnewline
$i$ & $-20.25_{-0.19}^{+0.18}$ & $0.54_{-0.18}^{+0.20}$ & $3.95\pm0.16$ & $4.33$ & $4.55_{-0.48}^{+0.48}$ \tabularnewline
$z$ & $-20.60_{-0.17}^{+0.18}$ & $0.36_{-0.15}^{+0.17}$ & $3.96\pm0.10$ & $3.79$ & $5.44_{-0.54}^{+0.54}$ \tabularnewline
$Y$ & $-20.57_{-0.18}^{+0.18}$ & $0.37_{-0.15}^{+0.17}$ & $3.72\pm0.13$ & $4.73$ & $4.95_{-0.49}^{+0.49}$ \tabularnewline
$J$ & $-20.97_{-0.17}^{+0.17}$ & $0.10_{-0.11}^{+0.14}$ & $3.59\pm0.41$ & $3.46$ & $6.19_{-0.74}^{+0.74}$ \tabularnewline
$H$ & $-21.32_{-0.18}^{+0.18}$ & $0.01_{-0.12}^{+0.13}$ & $3.61\pm0.23$ & $3.57$ & $9.38_{-1.02}^{+1.02}$ \tabularnewline
$K$ & $-21.23_{-0.16}^{+0.17}$ & $-0.11_{-0.10}^{+0.12}$ & $3.40\pm0.29$ & $3.35$ & $12.03_{-1.04}^{+1.04}$ \tabularnewline

\hline 
\end{tabular}
\par\end{centering}

\caption{\label{tab:lumfuncmultiSbc}As Table \ref{tab:lumfuncmultisingle}
but for S(B)ab-S(B)cd galaxies.}
\end{table}

\renewcommand{\arraystretch}{1}

\renewcommand{\arraystretch}{1.5}

\begin{table}
\begin{centering}
\begin{tabular}{cccccc}
\hline 
Band & $M^{*}$ & $\alpha$ & $\phi^{*}/10^{-3}$ & $\chi^{2}/\nu$ & $j/10^{7}$\tabularnewline
 & (mag) &  & ($\mathrm{mag}^{-1}\mathrm{Mpc}^{-3}$) &  & ($\mathrm{L}_{\odot}\mathrm{Mpc}^{-3}$)\tabularnewline
\hline 
$u$ & $-17.26_{-0.39}^{+0.33}$ & $-0.82_{-0.55}^{+0.61}$ & $10.74\pm2.81$ & $0.42$ & $2.84_{-0.74}^{+0.74}$ \tabularnewline
$g$ & $-18.25_{-0.27}^{+0.23}$ & $-0.70_{-0.29}^{+0.32}$ & $10.55\pm1.17$ & $1.72$ & $2.17_{-0.12}^{+0.12}$ \tabularnewline
$r$ & $-18.46_{-0.16}^{+0.16}$ & $-0.40_{-0.19}^{+0.22}$ & $11.56\pm0.50$ & $1.08$ & $1.91_{-0.06}^{+0.06}$ \tabularnewline
$i$ & $-19.07_{-0.24}^{+0.23}$ & $-0.80_{-0.23}^{+0.26}$ & $8.73\pm1.33$ & $2.09$ & $2.27_{-0.13}^{+0.13}$ \tabularnewline
$z$ & $-19.24_{-0.26}^{+0.24}$ & $-0.84_{-0.26}^{+0.29}$ & $7.97\pm1.38$ & $1.55$ & $2.40_{-0.15}^{+0.15}$ \tabularnewline
$Y$ & $-19.52_{-0.34}^{+0.29}$ & $-1.20_{-0.26}^{+0.27}$ & $5.26\pm1.86$ & $0.73$ & $2.54_{-0.30}^{+0.30}$ \tabularnewline
$J$ & $-19.54_{-0.39}^{+0.34}$ & $-1.17_{-0.29}^{+0.32}$ & $4.65\pm0.98$ & $1.66$ & $2.31_{-0.13}^{+0.13}$ \tabularnewline
$H$ & $-19.97_{-0.44}^{+0.36}$ & $-1.36_{-0.27}^{+0.31}$ & $3.66\pm1.50$ & $1.47$ & $3.82_{-0.60}^{+0.60}$ \tabularnewline
$K$ & $-19.59_{-0.54}^{+0.45}$ & $-1.31_{-0.34}^{+0.41}$ & $3.97\pm1.70$ & $2.52$ & $4.25_{-0.82}^{+0.82}$ \tabularnewline

\hline 
\end{tabular}
\par\end{centering}

\caption{\label{tab:lumfuncmultiSd}As Table \ref{tab:lumfuncmultisingle}
but for Sd-Irr galaxies.}
\end{table}

\renewcommand{\arraystretch}{1}
\setlength{\tabcolsep}{6pt}

The majority of the morphology sub-populations appear to be well fitted
by a single Schechter function, with reduced $\chi^{2}$ values typically
lying within the range $0.5<\chi^{2}/\nu<2$. The only notable exceptions
to this are for the LBS and the Sab-Scd populations. The knee of the
LBS population at most wavelengths lies outside the fitting limits,
beyond the faint-end limit, and so the overall fit may instead be
better suited by a single exponential function. Note in Table \ref{tab:lumfuncmultiLBS}
that the slope of the LBS population in the $K$ band is so poorly
constrained that the estimated luminosity density diverges, hence
no estimate of luminosity density is calculated. The faint-end limit
may also affect the Sd-Irr population, particularly in the estimation
of the faint end slope, as can be seen in Figure \ref{fig:lumfuncmultisingleellipses}.
The Sab-Scd population is well described by a single Schechter fit
for systems fainter than $M_{r}\sim-21$ mag, however; at brighter
magnitudes the Sab-Scd population departs from a single Schechter
form, with number counts at brighter magnitudes very closely matching
those of the S0-Sa population. We find that a double Schechter fit
to this Sab-Scd population is similarly recalcitrant, and so we elect
to maintain a single Schechter fit to the Sab-Scd population.

It is evident however that at wavelengths longer than the $z$ band
a single Schechter fit to the total GLF is a poor fit, reaching a
peak goodness of fit value of $\chi^{2}/\nu=4.31$ in the $H$ band.
This is as expected if one considers that the field galaxy LF is comprised
of an initial red spheroidal `bump' at bright magnitudes and then
a subsequent blue disk `bump' at fainter magnitudes, as can clearly
be seen in Figure \ref{fig:lumfuncmultisingle}, and noted in, e.g.,
\citet{Phillipps1995,Popesso2006,Loveday2012}. A single Schechter
function is unable to account for the intricacy in this distribution.

\begin{figure*}
\includegraphics[width=1\textwidth]{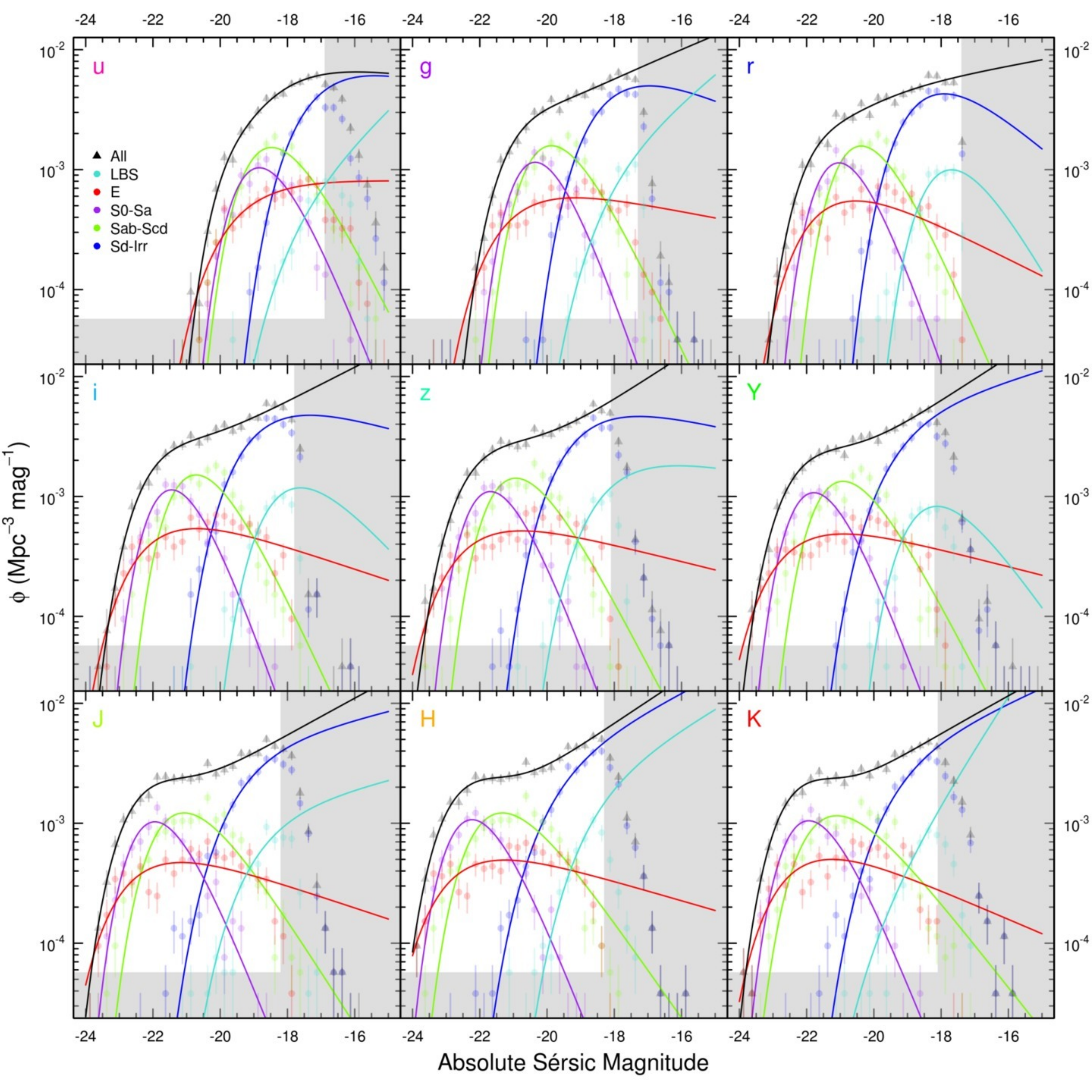}

\caption{\label{fig:lumfuncmultidouble}Morphological-type luminosity functions
across all nine bands for the various morphological types (coloured
points and lines, as indicated) and total populations (black points
and lines). Each morphological population has been fit with a single
Schechter function and is identical to those shown in Figure \ref{fig:lumfuncmultisingle}.
Total populations have been fit with a double Schechter function.
Prior to fitting, the data are split into bins of $0.25$ mag, with
the error on the measurement per bin taken as Poissonian ($\sqrt{n}$)
in nature. Shaded grey areas indicate those regions where data has
not been used in the fits. Variable faint-end magnitude limits are
given in Table \ref{tab:faintlimits}. The additional Schechter function
for the total population allows for the notable upturn at faint magnitudes
to be properly accounted for, especially at longer wavelengths.}
\end{figure*}

We elect to fit the total GLF with a double Schechter function with
a shared knee, while maintaining single Schechter fits to the morphology
sub-populations. The free parameters for the double Schechter fit
are $M^{*}$, $\alpha_{1}$, $\phi_{1}^{*}$, $\alpha_{2}$ and $\phi_{2}^{*}$.
The results of this fit are shown in Figure \ref{fig:lumfuncmultidouble}
for all nine bands, and the fit parameters given in Table \ref{tab:lumfuncmultidouble}.
It is instantly apparent once more that the GLF is more naturally
fit with a double Schechter function than a single Schechter function,
particularly so for the longer NIR wavelengths. All $\chi^{2}$ values
beyond the $z$ band show a significant improvement in the quality
of the fit. However, the shortest wavelengths show little need for
the extra parameters, with the goodness of fit showing a mild worsening
in the $u$ band, again most likely owing to the poorer quality of
the $u$ band data. Nevertheless, the overall fits appear robust,
and so we advocate a double Schechter form for the field total GLF
but single Schechter function forms for the morphology sub-population
MLFs.

\renewcommand{\arraystretch}{1.5}

\begin{table*}
\begin{centering}
\begin{tabular}{cccccccc}
\hline 
Band & $M^{*}$ & $\alpha_{1}$ & $\phi_{1}^{*}/10^{-3}$ & $\alpha_{2}$ & $\phi_{2}^{*}/10^{-3}$ & $\chi^{2}/\nu$ & $j/10^{7}$\tabularnewline
 & (mag) &  & ($\mathrm{mag}^{-1}\mathrm{Mpc}^{-3}$) &  & ($\mathrm{mag}^{-1}\mathrm{Mpc}^{-3}$) &  & ($\mathrm{L}_{\odot}\mathrm{Mpc}^{-3}$)\tabularnewline
\hline 
$u$ & $-18.53\pm1.25$ & $-0.91\pm7.17$ & $9.64\pm12.91$ & $1.25\pm6.01$ & $1.46\pm9.39$ & $1.23$ & $11.88_{-2.04}^{+2.04}$ \tabularnewline
$g$ & $-20.28\pm0.26$ & $-1.29\pm0.14$ & $3.51\pm1.75$ & $0.06\pm0.59$ & $4.88\pm1.22$ & $1.43$ & $14.16_{-0.58}^{+0.58}$ \tabularnewline
$r$ & $-20.90\pm0.26$ & $-1.13\pm0.07$ & $4.51\pm1.03$ & $0.53\pm0.56$ & $3.01\pm0.83$ & $2.71$ & $15.84_{-0.71}^{+0.71}$ \tabularnewline
$i$ & $-21.45\pm0.20$ & $-1.35\pm0.21$ & $2.20\pm1.50$ & $-0.09\pm0.48$ & $4.87\pm1.34$ & $1.41$ & $19.75_{-0.85}^{+0.85}$ \tabularnewline
$z$ & $-21.78\pm0.25$ & $-1.46\pm0.21$ & $1.40\pm1.12$ & $-0.26\pm0.44$ & $5.05\pm0.88$ & $1.25$ & $23.31_{-1.36}^{+1.36}$ \tabularnewline
$Y$ & $-21.76\pm0.24$ & $-1.45\pm0.23$ & $1.44\pm1.25$ & $-0.10\pm0.54$ & $4.83\pm0.84$ & $0.68$ & $22.67_{-1.09}^{+1.09}$ \tabularnewline
$J$ & $-21.82\pm0.17$ & $-1.38\pm0.13$ & $1.58\pm0.76$ & $0.08\pm0.35$ & $4.78\pm0.72$ & $1.27$ & $26.00_{-1.36}^{+1.36}$ \tabularnewline
$H$ & $-22.04\pm0.26$ & $-1.46\pm2.43$ & $1.35\pm6.33$ & $0.08\pm2.58$ & $5.30\pm6.69$ & $1.21$ & $38.78_{-3.01}^{+3.01}$ \tabularnewline
$K$ & $-21.72\pm0.23$ & $-1.39\pm1.62$ & $1.64\pm3.13$ & $0.24\pm1.55$ & $5.09\pm3.18$ & $0.86$ & $47.13_{-3.10}^{+3.10}$ \tabularnewline

\hline 
\end{tabular}
\par\end{centering}

\caption{\label{tab:lumfuncmultidouble}Double Schechter luminosity function
fit parameters for the total GLF as shown in Figure \ref{fig:lumfuncmultidouble}.
From left to right, columns are: GAMA passband; the shared knee in
the Schechter function ($M^{*}$); the primary slope of the faint
end of the Schechter function ($\alpha_{1}$); the primary normalisation
constant of the Schechter function ($\phi_{1}^{*}$); the secondary
slope of the faint end of the Schechter function ($\alpha_{2}$);
the secondary normalisation constant of the Schechter function ($\phi_{2}^{*}$);
the $\chi^{2}$ goodness of fit parameter ($\chi^{2}/\nu$), and;
the luminosity density ($j$). Errors are estimated from jackknifed
resampling using the relation $\sigma^{2}=\frac{N-1}{N}\sum_{i=1}^{N}\left(x_{j}-x\right)^{2}$,
where $x$ is the best fit parameter, $x_{j}$ is the best fit parameter
as given from a jackknife resampled variant of the data set and $N$
is the number of jackknife volumes. We adopt $N=10$.}
\end{table*}

\renewcommand{\arraystretch}{1}

\begin{figure*}
\includegraphics[width=1\textwidth]{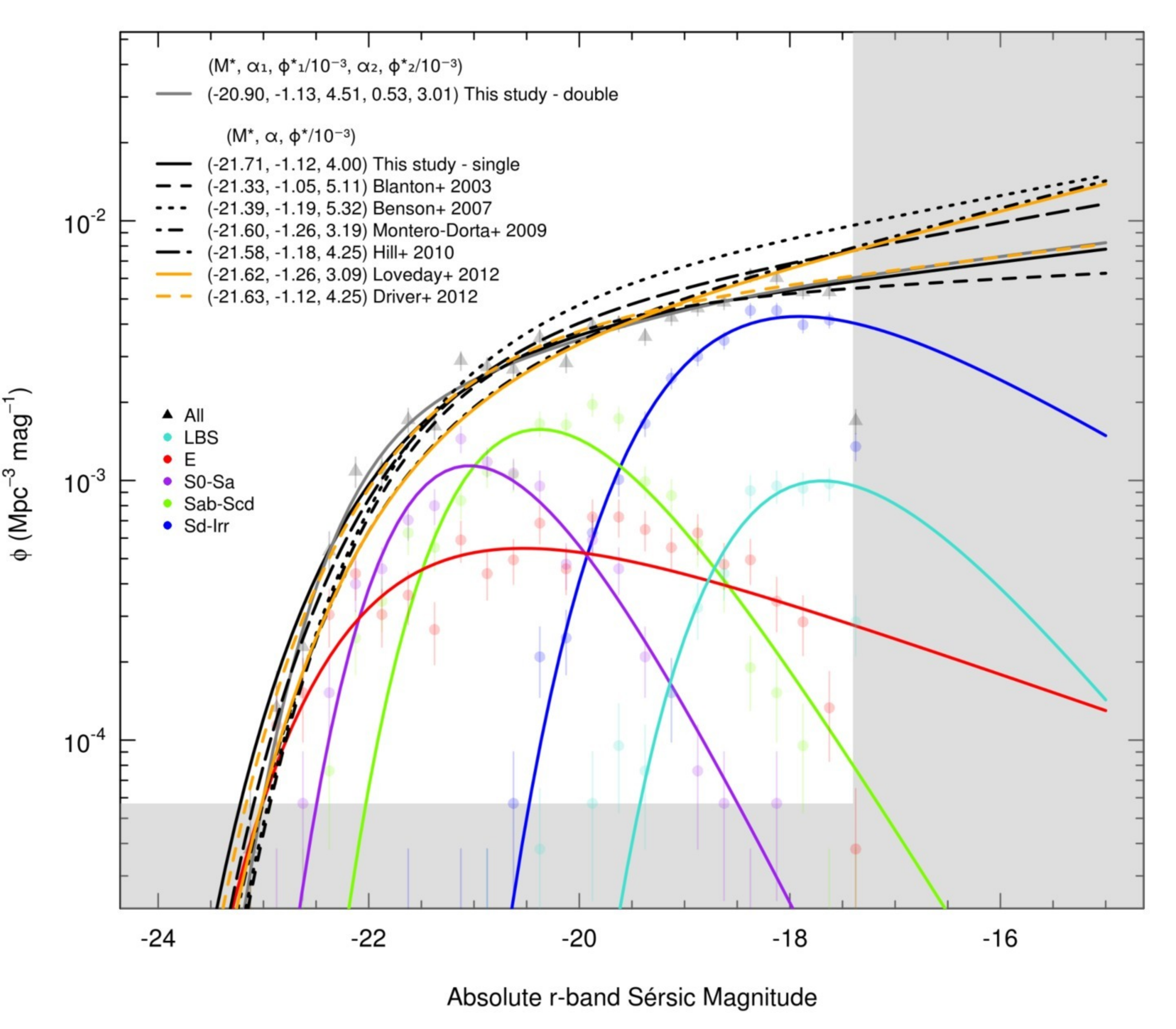}

\caption{\label{fig:lumfuncr}Morphological-type luminosity functions in the
$r$ band fit by single-Schechter functions in addition to the total
luminosity function fit by both a single and double-Schechter function,
shown in grey and black respectively. Each morphology is labelled
and coloured according to the inset legend. Prior to fitting, the
data are split into bins of $0.25$ mag, with the error per bin assumed
as Poissonian ($\sqrt{n}$) in nature. Shaded grey areas ($M>-17.4$
mag and $n\le3$) indicate those regions where data has not been used
in constraining the Schechter fits. Schechter fit parameters from
the global fits (inset, top left) in addition to single-Schechter
fits from other studies are also shown, for reference. Where appropriate,
Schechter fit data from other studies has been $k$-corrected from
$z=0.1$ back to a $z=0$ rest frame using a typical correction of
$k_{0.1}=0.12$. \citealp{Blanton2003a,Montero-Dorta2009,Loveday2012}
have been corrected in this fashion, whereas \citet{Benson2007,Hill2010,Driver2012}
have not. Note that the \citet{Benson2007} values have been scaled
up by a factor of $10$.}
\end{figure*}

A summary of both the single and double Schechter fits to the GLF
in addition to the adopted single Schechter fits to the MLFs in the
$r$ band are shown in Figure \ref{fig:lumfuncr}. Also shown are
several other contemporary single Schechter fits to similar $r$ band
data, scaled to our preferred cosmology of ($H_{0}$, $\Omega_{m}$,
$\Omega_{\Lambda}$) = ($70$, $0.3$, $0.7$) and $k$-corrected
where necessary from $r^{0.1}$ back to a $z=0$ rest frame using
a typical correction of $k_{0.1}=0.12$. There is generally good agreement
between our global luminosity function fits and those of other studies.
The variable faint end limit between surveys makes a comparison of
the faint end slope problematic, however, the $M^{*}$ and $\phi^{*}$
parameters agree well to within their errors. The need for a second
Schechter component in the $r$ band is less evident than at longer
wavelengths, however, its effects in causing a steeper drop off at
the bright end can clearly be seen in improving the fit to the data.

\section{The Cosmic Spectral Energy Distribution by Hubble Type}

\label{sec:csed}The morphological classifications derived in Section
\ref{sec:classification} are useful for many purposes beyond measuring
the luminosity distributions listed in Tables \ref{tab:lumfuncmultisingle}
to \ref{tab:lumfuncmultidouble}. One in particular is the subdivision
of the cosmic spectral energy distribution (CSED; \citealp{Hill2010})
\footnote{The CSED is distinct to the energy in photons within a fixed volume
(see for example \citealp{Dominguez2011}), as the majority of these
were generated at earlier epochs (i.e., the CSED is the instantaneous
energy production rate whereas the extragalactic background light
is the integrated energy production incorporating cosmological effects).%
} by morphological type. The CSED can be derived from the fitted luminosity
functions, or directly by summing the flux from a volume-limited galaxy
sample observed across a broad wavelength range (see \citealp{Driver2012}
for a discussion of the two methods, with the latter generally being
favoured if the data are sufficiently deep). The CSED describes the
instantaneous attenuated energy production rate of the Universe today.
The energy budget, like the mass budget, is a fundamental description
of the Universe which can be readily compared to complex (e.g., \citealp{Somerville2012})
or basic (e.g., \citealp{Driver2013}) model prescriptions.

At the present epoch the energy production budget is almost entirely
dominated and driven by stellar nucleosynthesis combined with dust
re-processing of the emergent starlight (i.e., the AGN contribution
at very low redshift is negligible, see \citealp{Driver2012}). Because
of this latter effect the CSED comes in two flavours, attenuated (i.e.,
as observed) and unattenuated (i.e., dust corrected), both of which
are useful. For example to measure the global star-formation rate
of a specific sub-population one desires the unattenuated CSED, but
to quantify the ambient intergalactic UV flux one requires the attenuated
CSED. 

While the attenuated CSED is straight-forward to derive, the unattenuated
CSED requires a correction for the wavelength and inclination dependent
photon escape fraction. The $FUV$ to $K$ photon-escape fraction,
integrated over all inclinations for the zero redshift galaxy population,
was recently quantified by \citet{Driver2007b,Driver2008} using data
from the Millennium Galaxy Catalogue (MGC; \citealp{Liske2003b,Driver2005}).
Here the MGC data was used to constrain the face-on central opacity
($\tau_{B}\sim3.8$) of the disk galaxy population \citep{Driver2007b}
by comparison of the inclination dependent $B_{{\rm MGC}}$-band $M^{*}$
value, with predictions from the detailed radiative transfer modelling
described in \citealp{Popescu2011} (see also \citealp{Popescu2000}
and \citealp{Tuffs2004}). It is worth noting that this dust prescription
incorporates full radiative transfer treatment including anisotropic
scattering processes from within three distinct dust components: an
extended optically thin double exponential dust disc, a compact optically
thick double exponential dust disc, and clumpy components associated
with star forming complexes, with a prescription that allows for cloud
fragmentation \citep{Tuffs2004}.

Here we report the contribution of each morphological type to both
the attenuated and unattenuated CSED using the photon escape fraction
described above for the S(B)ab-S(B)cd, Sd-Irr, and LBS populations,
and assuming the E and S(B)0-S(B)a populations are dust free. We opt
to dust correct LBS galaxies after a non-exhaustive examination of
the spectra of a large number of these systems wherein we found repeated
evidence for ongoing star formation. While Sa galaxies in our sample
may indeed contain dust, we assume that to first order a correction
of this type is broadly correct. Note that \citet{Rowlands2012} recently
showed from Herschel-ATLAS data that at most $10\%$ of the elliptical
population contains dust (see also \citealp{Agius2013}). In due course
the variation of dust properties with morphological type will be investigated
using the FUV to far-IR GAMA multi-wavelength dataset (see Driver
et al., in prep).

\subsection{Measuring the integrated fluxes}

Any description of the CSED will be incomplete without the inclusion
of FUV and NUV estimates. This is because almost $40\%$ of the energy
of a global population emerges at wavelengths below $400$nm \citep{Driver2012}.
Rather than computing the full Sérsic luminosity functions as we have
done in the $ugrizYJHK$ bands, here we simply elect to sum the FUV
and NUV flux for the distinct samples directly and divide by the volume
probed. Our FUV and NUV data are taken from the GALEX satellite, specifically;
a combination of Medium Imaging Survey (MIS) archival and proprietary
data obtained by the MIS and GAMA teams (see \citealp{Driver2012}
for further details). 

Table \ref{tab:lds} shows the luminosity density values derived directly
by summing the fluxes of all systems within our volume and for each
population. Only galaxies which lie in the common region (i.e., sampled
by all 11 bands, see \citealp{Driver2012}) are included and the volume
is modified accordingly to compensate ($\times0.86$). The luminosity
densities shown in Table \ref{tab:lds} can be compared to those derived
from the fitted Schechter functions in Tables \ref{tab:lumfuncmultisingle}
to \ref{tab:lumfuncmultidouble}. As discussed in \citet{Driver2012},
discrepancies between these two estimates can arise from the extrapolation
of the fitted Schechter function combined with sub-optimal fits around
the $L^{*}$-region. The sum of these values for the individual morphological
classes also agree well with the global values reported in \citet{Driver2012},
implying internal consistency between the various GAMA sub-samples
and methodologies. In detail the FUV and NUV values reported here
are lower which is also consistent with the slightly lower median
redshift given the steeply declining cosmic star-formation history
(see for example \citealp{Hopkins2006,Driver2013}) --- i.e., $z\sim0.04$
versus $z\sim0.08$, equivalent to a time interval of $\sim0.5$ Gyr.
Note, from \citet{Driver2013}, we expect the mean cosmic star formation
rate at $z=0.06$ to be $\sim22\%$ higher than at $z=0$.

\renewcommand{\arraystretch}{1.5}
\setlength{\tabcolsep}{6pt}

\begin{table*}
\begin{tabular}{c|c|c|c|c|c|c|c} \hline
\multicolumn{2}{c}{Wavelength} & \multicolumn{6}{c}{Hubble type} \\ \cline{3-8}
band & ($\mu$m) & All & E & S(B)0-S(B)a & S(B)ab-S(B)cd & Sd-Irr & LBS  \\ \cline{3-8}
  &       & \multicolumn{6}{c}{$10^7 \mathrm{L}_{\odot}\mathrm{Mpc}^{-3}$} \\ \hline
FUV & 0.153  & 16378.14  $\pm$  2504.93  & 699.05  $\pm$  106.91  & 1928.87  $\pm$  295.01  &  7069.65  $\pm$  1081.25  & 3010.56  $\pm$  460.44  & 380.42  $\pm$   58.18  \\
NUV & 0.230  &  115.64  $\pm$   17.69  &   7.30  $\pm$    1.12  &   15.56  $\pm$    2.38  &   49.45  $\pm$    7.56  &  19.95  $\pm$    3.05  &   2.48  $\pm$    0.38  \\
u & 0.355  &   11.93  $\pm$    0.95  &   1.95  $\pm$    0.16  &    2.53  $\pm$    0.20  &    4.44  $\pm$    0.36  &   1.45  $\pm$    0.12  &   0.17  $\pm$    0.01  \\
g & 0.467  &   12.56  $\pm$    1.01  &   2.78  $\pm$    0.22  &    3.30  $\pm$    0.26  &    4.11  $\pm$    0.33  &   1.25  $\pm$    0.10  &   0.14  $\pm$    0.01  \\
r & 0.616  &   15.13  $\pm$    1.21  &   3.77  $\pm$    0.30  &    4.32  $\pm$    0.35  &    4.64  $\pm$    0.37  &   1.33  $\pm$    0.11  &   0.15  $\pm$    0.01  \\
i & 0.747  &   18.05  $\pm$    1.44  &   4.68  $\pm$    0.37  &    5.32  $\pm$    0.43  &    5.45  $\pm$    0.44  &   1.49  $\pm$    0.12  &   0.16  $\pm$    0.01  \\
z & 0.892  &   21.54  $\pm$    1.72  &   5.83  $\pm$    0.47  &    6.66  $\pm$    0.53  &    6.22  $\pm$    0.50  &   1.61  $\pm$    0.13  &   0.18  $\pm$    0.01  \\
Y & 1.030  &   22.06  $\pm$    1.76  &   6.05  $\pm$    0.48  &    7.08  $\pm$    0.57  &    6.24  $\pm$    0.50  &   1.56  $\pm$    0.12  &   0.18  $\pm$    0.01  \\
J & 1.248  &   25.80  $\pm$    2.06  &   7.10  $\pm$    0.57  &    8.56  $\pm$    0.68  &    7.32  $\pm$    0.59  &   1.61  $\pm$    0.13  &   0.20  $\pm$    0.02  \\
H & 1.631  &   38.41  $\pm$    3.07  &  10.65  $\pm$    0.85  &   12.88  $\pm$    1.03  &   10.80  $\pm$    0.86  &   2.46  $\pm$    0.20  &   0.27  $\pm$    0.02  \\
K & 2.201  &   45.60  $\pm$    3.65  &  12.47  $\pm$    1.00  &   15.40  $\pm$    1.23  &   12.95  $\pm$    1.04  &   2.88  $\pm$    0.23  &   0.32  $\pm$    0.03  \\ \hline
\end{tabular}

\caption{\label{tab:lds}Luminosity densities as a function of wavelength and
morphological type.}
\end{table*}

\renewcommand{\arraystretch}{1}

\subsection{Star-formation rates by morphological type}

The dust corrected FUV luminosity density can be converted directly
to a measure of the star-formation rate. These values are shown in
Table \ref{tab:uvdata} and assume a photon escape fraction of $23\%$.
In brief, this photon escape fraction is determined by deriving the
galaxy luminosity function in the $B$ band for galaxies taken from
the Millennium Galaxy Catalogue \citep{Liske2003b}, subdivided by
inclination. The trends in $M^{*}$ with $\cos(i)$ are compared to
those predicted by the complex dust models of \citet{Tuffs2004} (see
also \citealp{Popescu2011}), and used to constrain the face on central
opacity \citep{Driver2007b}. In \citet{Driver2008}, this value is
used to predict the average photon escape fraction as a function of
wavelength (see Table $3$ of \citealp{Driver2012}). We use the standard
prescription by \citet{Kennicutt1989} to derive the star-formation
rate which is based on a \citet{Salpeter1955} initial mass-function.
The values reported in Table \ref{tab:uvdata} are typically $\times2$
higher than those reported by \citet[see their Figure 7]{James2008}.
As no evolutionary corrections to the magnitudes are applied, these
values correspond to a measurement of the star-formation rate at the
median redshift of $z\sim0.04$ (again, a $\sim0.5$Gyr time interval).
Additionally, our `All' measurement is approximately two times lower
than that reported in \citet{Robotham2011a}, which is again consistent
when taking into account the median redshift offset between these
two datasets. Our data also confirm the trend seen by \citet{James2008},
that the SFR density in the nearby Universe is dominated by the intermediate
S(B)ab-S(B)cd Hubble types, with a sharp decline towards earlier or
later types. 

\renewcommand{\arraystretch}{1.5}
\setlength{\tabcolsep}{1pt}

\begin{table}
\begin{center}
\begin{tabular}{c|c|c} \hline
Hubble & $SFR$ & $SFR$ Contribution \\ 
Type            & (M$_{\odot}$ yr$^{-1}$ Mpc$^{-3}$) & $\%$ \\ \hline
 All            & $0.0125 \pm 0.0030$ & $100$ \\
 E              & $0.0001 \pm 0.0001$ & $1$ \\
 S(B)0-S(B)a    & $0.0020 \pm 0.0003$ & $16$ \\
 S(B)ab-S(B)cd  & $0.0073 \pm 0.0011$ & $58$ \\
 Sd-Irr         & $0.0031 \pm 0.0005$ & $25$ \\
 LBS            & $<0.0001 \pm 0.0001$ & $<1$ \\ \hline
\end{tabular}
\end{center}

\caption{\label{tab:uvdata}The star-formation rate density for each morphological
type as derived from the FUV luminosity densities reported in Table
\ref{tab:lds}.}
\end{table}

\renewcommand{\arraystretch}{1}

\subsection{The attenuated CSED}

\begin{figure}
\includegraphics[width=1\columnwidth]{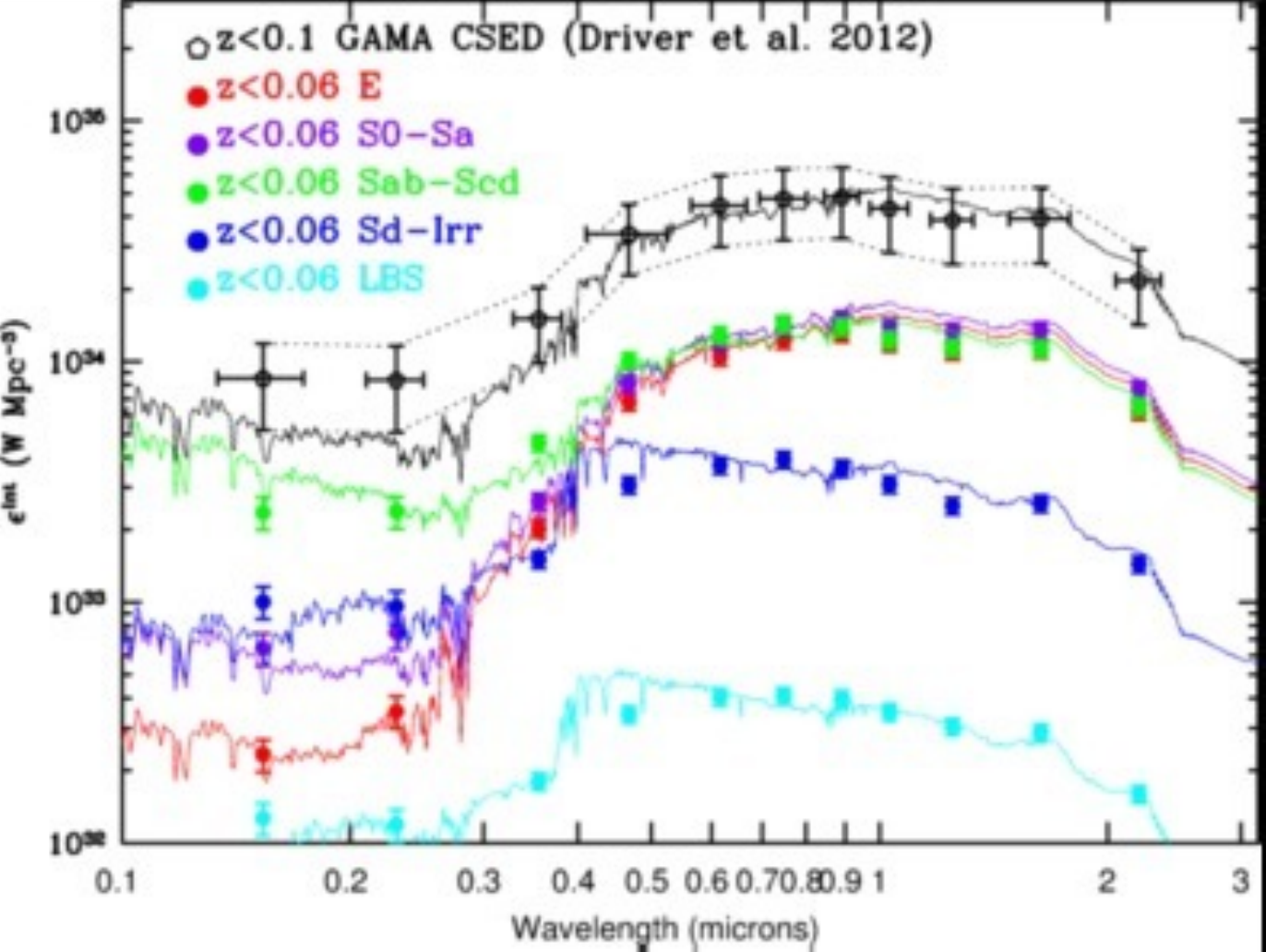}

\caption{\label{fig:acsed}The attenuated (as observed) CSED. Integration under
each line provides a direct measure of the emergent observed instantaneous
energy production for each galaxy population.}

\end{figure}

Figure \ref{fig:acsed} shows the full $FUV$ to $K$ attenuated CSED
(i.e., as observed) for each of the populations. Overlaid as black
data points is the global CSED reported in \citet{Driver2012} derived
for the full GAMA $z<0.1$ sample. Note the earlier data include the
sample variance uncertainty indicated by the error bars and dotted
uncertainty ranges. As here we are interested in the variations between
the morphological types within a single volume we do not include the
cosmic variance errors. For each morphological type we fit a range
of single stellar population (SSP) \noun{PEGASE} models (see \citealp{Fioc1999})
to our data. The best fits are shown in Figure \ref{fig:acsed} by
the colour indicated in the key. The sum of these fits is shown as
the black curve which agrees well with the $z<0.1$ CSED showing only
a slight discrepancy in the UV where one might expect a slightly reduced
CSED due to the declining star-formation rate from $z=0.08$ to $z=0.04$
(i.e., $\Delta t\sim0.5$ Gyr). We do not report the \noun{PEGASE}
values for these curves as they are simply being used here as appropriate
fitting functions. Integrating these functions therefore provides
a direct measure of the instantaneous energy production emerging from
each galaxy population. The total energy output is $(8.53\pm0.20)\times10^{34}$
W Mpc$^{-3}$ with approximate subdivisions of $27$ per cent, $31$
per cent, $32$ per cent, $9$ per cent, and $1$ per cent arising
from the E, S(B)0-S(B)a, S(B)ab-S(B)cd, Sd-Irr and LBS populations
respectively. This is surprisingly well balanced and shows that all
galaxy types contribute significantly to the ambient inter-galactic
radiation field, i.e., $\sim58$ per cent spheroid-dominated and $\sim42$
per cent disk-dominated.

\subsection{The unattenuated CSED}

\begin{figure}
\includegraphics[width=1\columnwidth]{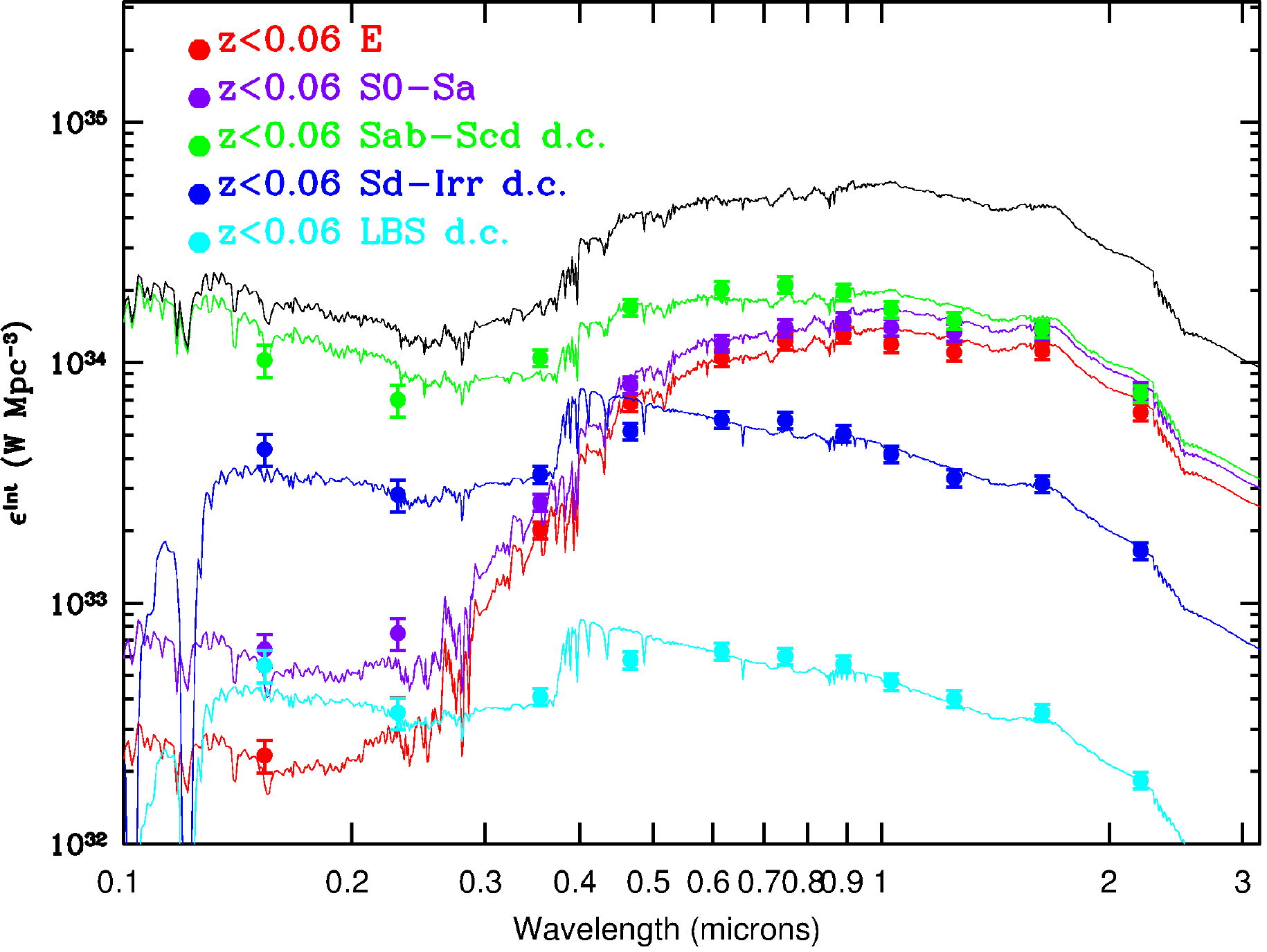}

\caption{\label{fig:ccsed}The unattenuated (corrected) CSED. Integration under
each line provides a direct measure of the emergent corrected instantaneous
energy production for each galaxy population.}

\end{figure}

Figure \ref{fig:ccsed} shows the unattenuated (corrected) CSED for
each of the populations by applying the photon escape fraction prescription
determined in \citet{Driver2008} to the S(B)ab-S(B)cd, Sd-Irr, and
LBS populations only. Similarly these data are fitted to a range of
\noun{PEGASE} SSP model as before and integrated to give the instantaneous
energy production of $(1.12\pm0.15)\times10^{35}$ W Mpc$^{-3}$ approximately
subdivided by $21$ per cent, $23$ per cent, $44$ per cent, $11$
per cent and $1$ per cent for the E, S(B)0-S(B)a, S(B)ab-S(B)cd,
Sd-Irr and LBS populations respectively. Hence we see that although
the energy which enters into the IGM is dominated $58$:$42$ by spheroid-dominated:disk-dominated
types the actual energy production rate is almost inverted, i.e.,
$44$:$56$ spheroid-dominated:disk-dominated.

\section{Conclusions}

We have defined a local ($0.025<z<0.06$) volume limited sample of
$3,727$ galaxies (GAMAnear) taken from the Galaxy And Mass Assembly
survey (GAMA; \citealp{Driver2009}). Using GAMA-reprocessed $9$
band imaging from the SDSS ($ugriz$; \citealt{York2000,Abazajian2009})
and UKIDSS-LAS ($YJHK$; \citealp{Lawrence2007}) surveys, we provide
robust visual classifications for each galaxy in our sample into its
morphological Hubble type (E, S(B)0-S(B)a, S(B)ab-S(B)cd, Sd-Irr),
alongside an additional `little blue spheroid' (LBS) class; a class
of potential blue dwarf ellipticals, and a contaminant `Star' class.
Approximately $27$\% of this sample is visually classified as spheroid
dominated, with the remaining $\sim73$\% visually classified as disk
dominated or LBS. We explore morphological trends with several global
measurements, namely; $r$ band half-light radius (kpc); ellipticity
as measured in the $r$ band, absolute $r$ band Sérsic magnitude
(truncated at $10$ $r_{e}$); rest-frame $(u-r)$ colour, and; $r$
band Sérsic index. In these global parameter spaces, we are able to
reproduce several well known morphological relations, including the
curved magnitude-radius relation for elliptical galaxies \citep{Graham2008b,Forbes2008}
and the galaxy population bimodality as has been shown in, e.g., \citet{Baldry2004,Driver2006,Kelvin2012}.

Using GAMA single-Sérsic \citep{Sersic1963,Sersic1968} structural
measurements (\citealp{Kelvin2012}), we maintain that the most meaningful
measurement of the total flux of a galaxy is that given by the Sérsic
magnitude, truncated at $10$ multiples of the half-light radius.
This estimate of total flux allows us to derive luminosity functions
for both the global population and the constituent morphology sub-populations
in each passband ($ugrizYJHK$). We confirm that the total galaxy
luminosity function (GLF) is best described by a double-Schechter
form \citep{Schechter1976} with a single distinctive `knee' ($L^{*}$/$M^{*}$)
parameter. Conversely, we find the constituent morphological-type
luminosity functions (MLFs) are well described by a single-Schechter
form. Tables \ref{tab:lumfuncmultisingle} to \ref{tab:lumfuncmultidouble}
provide full Schechter fit parameters for these data across all $9$
wavelengths.

Our morphological classifications allow for the division of the cosmic
spectral energy distribution (CSED) by morphological type. The CSED
describes the instantaneous energy production rate of the Universe,
providing a means by which cosmological model predictions of the total
local energy budget may be tested. Here we provide estimates of the
CSED by directly summing the flux in our volume limited sample for
each morphological type across each wavelength. Note that we include
flux measurements from the $FUV$ and $NUV$ in order to account for
the significant energy contribution at wavelengths below $400$ nm.
The energy production budget today is mainly comprised of both stellar
nucleosynthesis and dust reprocessing. Therefore, we have measured
both the attenuated (i.e., as observed) and unattenuated (i.e., dust
corrected) CSEDs for each morphological type by fitting a series of
single stellar population PEGASE models \citep{Fioc1999} to each
population. 

Based on our dust corrected $FUV$ flux estimates, we also construct
estimates of the local ($z\sim0.04$) star formation rate (SFR) density
subdivided by morphology. We find the SFR density across this redshift
range to be dominated by the intermediate S(B)ab-S(B)cd morphological
type systems, declining sharply at earlier or later Hubble types,
and confirming the trend seen in \citet{James2008}.

In addition, we find that $\sim58$\% of the total attenuated (observed)
energy output in the local Universe emerges from spheroid dominated
galaxies, with the remaining $\sim42$\% found in disk dominated systems.
The summation of these fits gives a total observed energy output of
$(8.53\pm0.20)\times10^{34}$ W Mpc$^{-3}$, in good agreement with
that of \citet{Driver2012}. The unattenuated CSEDs are derived by
applying the photon escape fraction prescription detailed in \citet{Driver2008},
calibrated using the radiative transfer models of \citet{Popescu2011},
to the S(B)ab-S(B)cd, Sd-Irr and LBS populations only. We find that
$\sim44$\% of the total unattenuated (corrected) energy output in
the local Universe emerges from spheroid dominated galaxies, with
the remaining $\sim56$\% found in disk dominated systems. The summation
of these fits gives a total corrected energy output in the local Universe
of $(1.12\pm0.15)\times10^{35}$ W Mpc$^{-3}$.

\section*{Acknowledgements}

This work was supported by the Austrian Science Foundation FWF under
grant P23946. AWG was supported under the Australian Research Council\textquoteright{}s
funding scheme FT110100263. GAMA is a joint European-Australasian
project based around a spectroscopic campaign using the Anglo-Australian
Telescope. The GAMA input catalogue is based on data taken from the
Sloan Digital Sky Survey and the UKIRT Infrared Deep Sky Survey. Complementary
imaging of the GAMA regions is being obtained by a number of independent
survey programs including GALEX MIS, VST KiDS, VISTA VIKING, WISE,
Herschel-ATLAS, GMRT and ASKAP providing UV to radio coverage. GAMA
is funded by the STFC (UK), the ARC (Australia), the AAO, and the
participating institutions. The GAMA website is http://www.gama-survey.org/.

\bibliographystyle{mn2e}
\bibliography{biblib}

\appendix

\section{Redshift Limits}

\label{app:redshiftlimits}Based on prior measurements of structural
properties as presented in \citet{Allen2006} and \citet{Simard2011},
we calculate the typical physical sizes of bulges and disks in the
local Universe. Adopting the redshifts provided in each respective
catalogue, we convert their reported bulge and disk angular sizes
to physical sizes (in kpc) in accordance with our preferred cosmology,
namely: ($H_{0}$, $\Omega_{m}$, $\Omega_{\Lambda}$)$=$($70$ km
s$^{-1}$ Mpc$^{-1}$, $0.3$, $0.7$). An appropriate conversion
between exponential disk scalelength (as provided in both catalogues)
and half-light radius is achieved using the well known relation
\[
r_{e}=b^{n}h
\]
where $h$ is the disk scalelength, $n$ is the Sérsic index (a measure
of the shape of the galaxy light profile; see Section \ref{sub:abssersicmags}
for further details) and $b$ is a function of $n$. For $n=1$, $b^{n}=1.678$.
We discard those model fits that lie outwith the range $0.1<B/T<0.9$,
limiting each catalogue to those systems that are not dominated by
the flux from a single component. \citet{Allen2006} model fits are
to $B$-band imaging data, whereas \citet{Simard2011} fits are in
the SDSS $r$-band. As shown in \citet{Kelvin2012}, one expects a
variation in observed half-light radii with wavelength. The best fitting
relations for both spheroidal (bulge) and disk components are given
by:

\begin{equation}
\log r_{\mathrm{e,sph}}=-0.304\log\lambda_{\mathrm{rest}}+1.506\label{eq:sphre}
\end{equation}
\begin{equation}
\log r_{\mathrm{e,disk}}=-0.189\log\lambda_{\mathrm{rest}}+1.176\label{eq:diskre}
\end{equation}
where $\lambda_{\mathrm{rest}}$ is the rest-frame wavelength. Accordingly,
we correct the \citet{Allen2006} half-light radii from $B$-band
($445$ nm) to the $r$-band ($622$ nm). We match both catalogues
to the GAMA-I tiling catalogue (version $16$) to limit our analyses
to galaxies that lie within the GAMA volume, and calculate $3$-sigma-clipped
robust mean values for the bulge and disk components in both studies. 

\begin{figure}
\includegraphics[width=1\columnwidth]{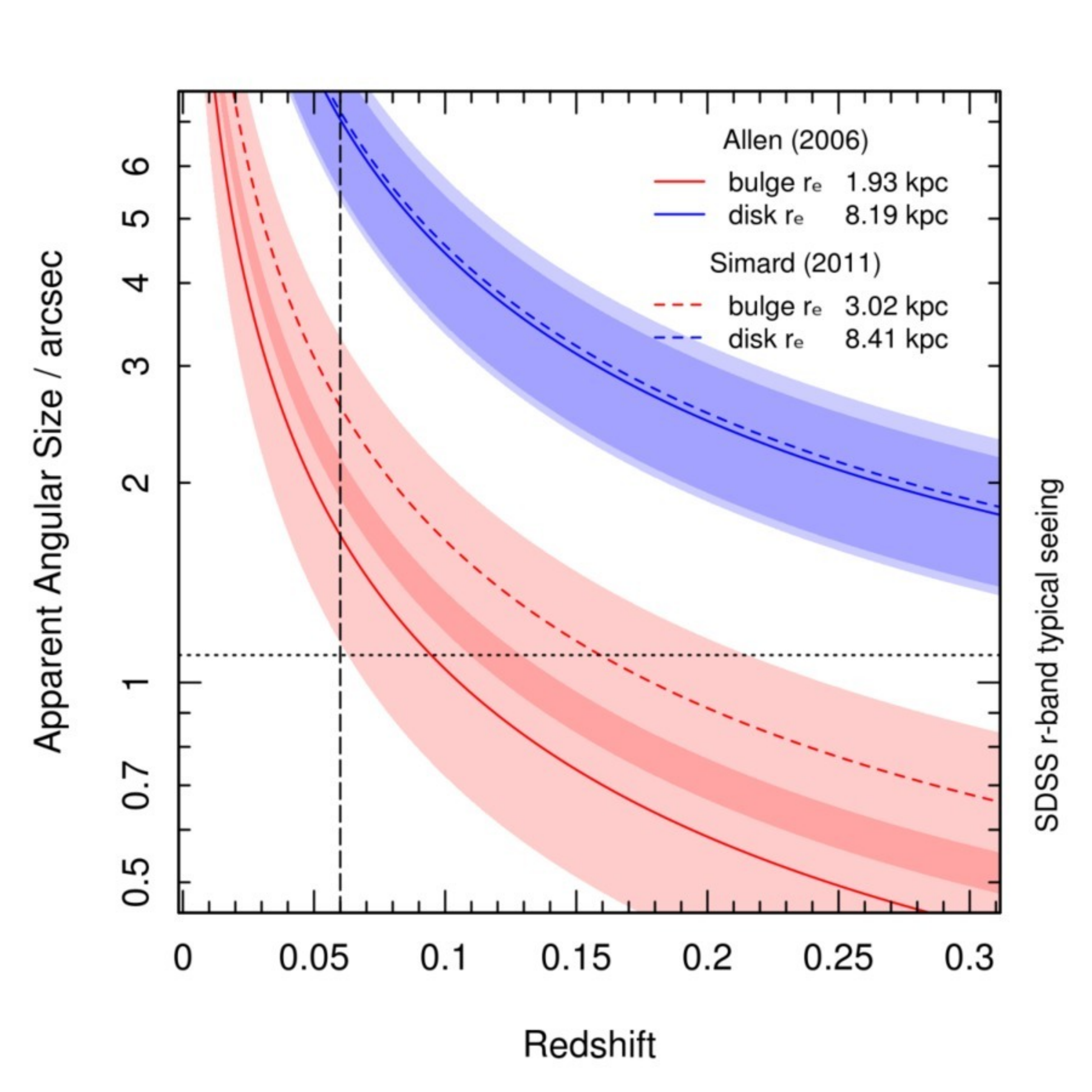}

\caption{\label{fig:angsize}Apparent angular size for typical bulges and disks
at varying redshifts. The red (blue) solid (dashed) line represents
the apparent angular size for bulges (disks) in the Allen (Simard)
data, as indicated. The shaded semi-transparent regions around each
line represents the half-sigma scatter in the data. In addition, the
horizontal dotted line lies at an angular size of $1.1$'', which
corresponds to the typical $r$-band seeing in SDSS \citep{Kelvin2012}.
Our chosen upper redshift limit of $z=0.06$ is shown as a vertical
dashed line.}
\end{figure}

We find the typical sizes for bulge components in the local Universe
as measured in the $r$-band to be $1.93\pm1.20$ kpc and $3.02\pm1.65$
kpc for Allen and Simard, respectively. We find the typical sizes
for the corresponding disk components to be $8.19\pm3.62$ kpc and
$8.41\pm4.45$ kpc, for Allen and Simard, respectively%
\footnote{Although it is crucial for us to estimate the typical \emph{observed}
sizes of bulges and disks in the local Universe when defining appropriate
sample redshift limits, we note that due to the effects of dust, projection
effects and bulge/disk decomposition considerations, the measured
sizes differ from the intrinsic underlying ones. Using the corrections
from \citet{Pastrav2013a,Pastrav2013b} for $\tau_{B}^{f}=3.8$ (the
same average dust opacity used to correct for dust attenuation), we
obtain average intrinsic bulge sizes of $1.80$ kpc and $2.82$ kpc
for the Allen and Simard samples, respectively, with corresponding
intrinsic disk sizes of $7.00$ kpc and $6.19$ kpc.%
}. Figure \ref{fig:angsize} shows the apparent angular size for structures
of these physical sizes at varying redshift. The red (blue) solid
(dashed) line represents the apparent angular size for bulges (disks)
in the Allen (Simard) data, as indicated. The shaded semi-transparent
regions around each line represents the half-sigma scatter in the
data. In addition, the horizontal dotted line lies at an angular size
of $1.1$'', which corresponds to the typical $r$-band seeing in
SDSS \citep{Kelvin2012}. 

For this study, we define an upper redshift limit of $z=0.06$. This
limit is chosen such that the majority of bulges (the limiting structural
component) should remain resolvable. A lower limit of $z=0.025$ is
also adopted to avoid low galaxy number densities below this redshift
and to ensure that measured redshifts are not dominated by peculiar
velocities.

\section{Comparison with Galaxy Zoo Morphologies}

\label{app:galaxyzoocomparison}To test our visual classifications
we compare our morphological classifications to those of the well-established
citizen science project Galaxy Zoo \citep{Lintott2008}. We employ
the Galaxy Zoo 1 data release (GZ1; \citealp{Lintott2011}, Table
2) in our analysis below.

GZ1 contains $667,944$ sources down to an SDSS apparent magnitude
limit of $r=17.77$ mag for all galaxies in the SDSS Data Release
7 which have spectra included. Of these $667,944$ objects, $1,779$
galaxies exhibit a direct match with the galaxies in our volume limited
GAMAnear sample of $3,727$ ($\sim48$\%) when matching by SDSS object
ID (OBJID). Each galaxy in the Galaxy Zoo catalogue is classified
as either `elliptical', `spiral' or `uncertain', with an associated
probability. We adopt a probability threshold of $80\%$. A markedly
high fraction of this subset is classified by Galaxy Zoo as `uncertain'
($1,050$; $59.0$\%), with the remainder as `elliptical' ($143$;
$8.0$\%) and `spiral' ($586$; $32.9$\%). For comparison, our own
matched sub-sample is similarly split into three comparable classification
bins: elliptical ($324$; $18.2\%$), spiral%
\footnote{Although lenticular and irregular types exist within this combined
population, we label it `spiral' for brevity and ease of comparison
to the Galaxy Zoo data.%
} (S0-Sa$\rightarrow$Sd-Irr, $1416$; $79.6\%$) and LBS/star ($39$;
$2.2\%$).

\begin{figure*}
\includegraphics[width=1\textwidth]{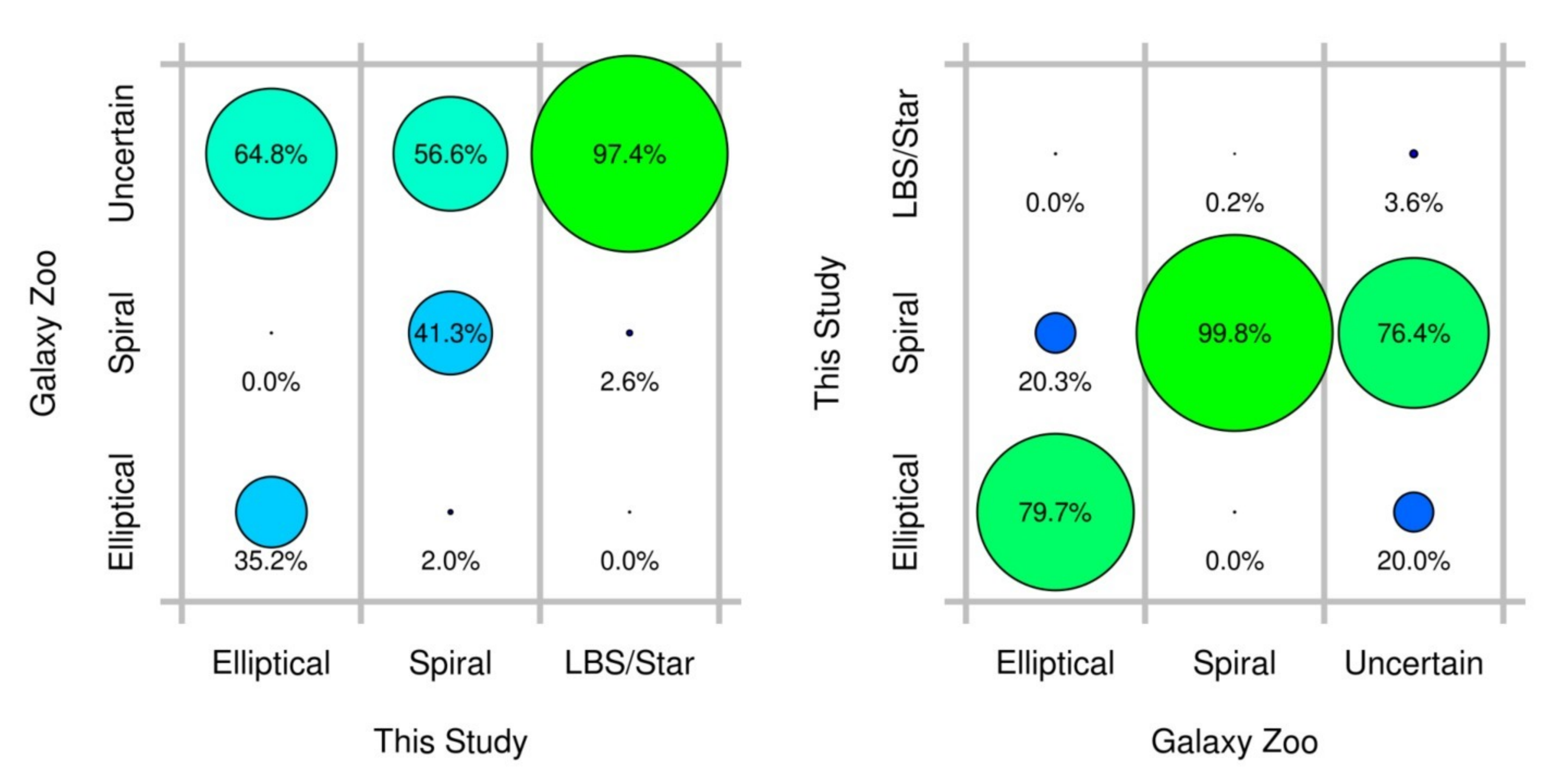}

\caption{\label{fig:galaxyzoo}A visual representation of the level of agreement
between our visual classifications and those provided by the Galaxy
Zoo project \citep{Lintott2008,Lintott2011}. These figures are constructed
using a common matched dataset of $1,779$ galaxies from GAMAnear
and Table 2 of \citet{Lintott2011}. Percentages shown depict the
fractional agreement with our own classifications (left) and with
the Galaxy Zoo classifications (right), that is; percentages in any
given column total $100\%$.}
\end{figure*}

Figure \ref{fig:galaxyzoo} shows the cross-correlation results between
our own visual classifications and those provided by Galaxy Zoo. The
number of galaxies within each bin are shown as `correlation bubbles',
with larger bubbles corresponding to a higher fraction of objects
within that bin. The fraction of galaxies within each classification
bin is quantified as a percentage of galaxies in our own study (left)
and of galaxies from Galaxy Zoo (right). As is shown, the vast majority
($99.8$\%) of the Galaxy Zoo spiral population are also classified
as spirals by our method (i.e., the Galaxy Zoo spiral population is
essentially a subset of our own), but not all of our spiral galaxies
are found to be spiral in the Galaxy Zoo data. A similarly large fraction
of the Galaxy Zoo elliptical population ($79.7$\%) are also classified
as elliptical by our method, with approximately one fifth of Galaxy
Zoo ellipticals classified as spirals in this study. The majority
of Galaxy Zoo uncertain galaxies are classified as spiral by our method
($76.4$\%), which may be expected owing to the typically fainter
surface brightnesses of galaxies of this type. As is shown in the
left panel, more than half of each grouping fall within the Galaxy
Zoo uncertain class, almost accounting for the entirety of our LBS/star
population. This large Galaxy Zoo uncertain population no doubt arises
due to the stringent $80$\% classification criterion recommended
for use in \citet{Lintott2011}. We note that the application of a
lower $66\%$ threshold (in line with our own classification method)
forces a significant fraction of the uncertain population into the
two standard `elliptical' and `spiral' sub-populations, in good agreement
with our own results, albeit with a larger margin of error. Despite
the large fraction of uncertain galaxies, we opt to maintain the recommended
classification criterion of $80$\% for our analyses. If one removes
the uncertain grouping from this figure, we find that the primary
axis (i.e., the {[}Elliptical,Elliptical{]}$\rightarrow${[}Spiral,Spiral{]}
axis) remains strong when using either our own method or Galaxy Zoo
as a reference baseline, indicating a good level of agreement between
our own classifications and those of Galaxy Zoo.

To summarise, while we acknowledge that Galaxy Zoo morphologies are
preferred for studies that require robust morphological information
for a large ($>10^{5}$) number of systems, we advocate that detailed
visual inspection by a team of experts produces notable advantages
over Galaxy Zoo for small datasets such as that presented in this
study. The creation of our own classification schema has allowed us
full control over, for example: classification criteria (and therefore
the ultimate resolution on available Hubble types); postage stamp
image creation, including red-green-blue filter selection, image sizes
and image scaling (both logarithmic and arctan), and, significantly;
our final sample selection. As shown in Figure \ref{fig:galaxyzoo},
we find a good level of agreement in morphological type between those
galaxies that exist in both the Galaxy Zoo dataset and our own GAMAnear
sample. This confirms that our classification schema is robust and
equally applicable to those additional galaxies in our sample that
do not have a counterpart in the Galaxy Zoo database.

\section{Morphologies in Colour-Index Space}

\label{app:morphologies}
\begin{figure*}
\includegraphics[width=0.9\textwidth]{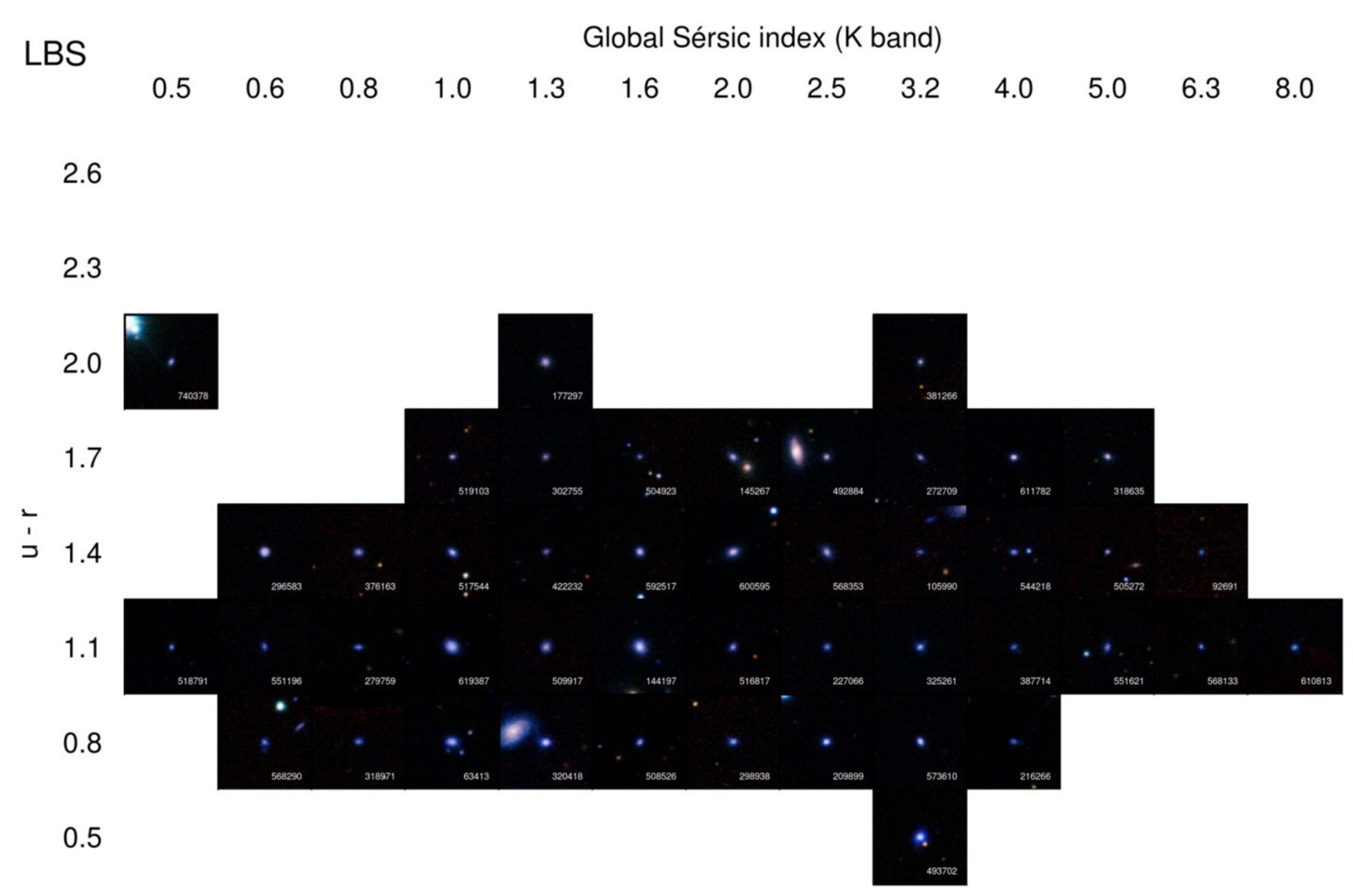}

\caption{\label{fig:eyeexam_LBS}Little Blue Spheroids in $u-r$ colour--Sérsic
index space.}
\end{figure*}

\begin{figure*}
\includegraphics[width=0.9\textwidth]{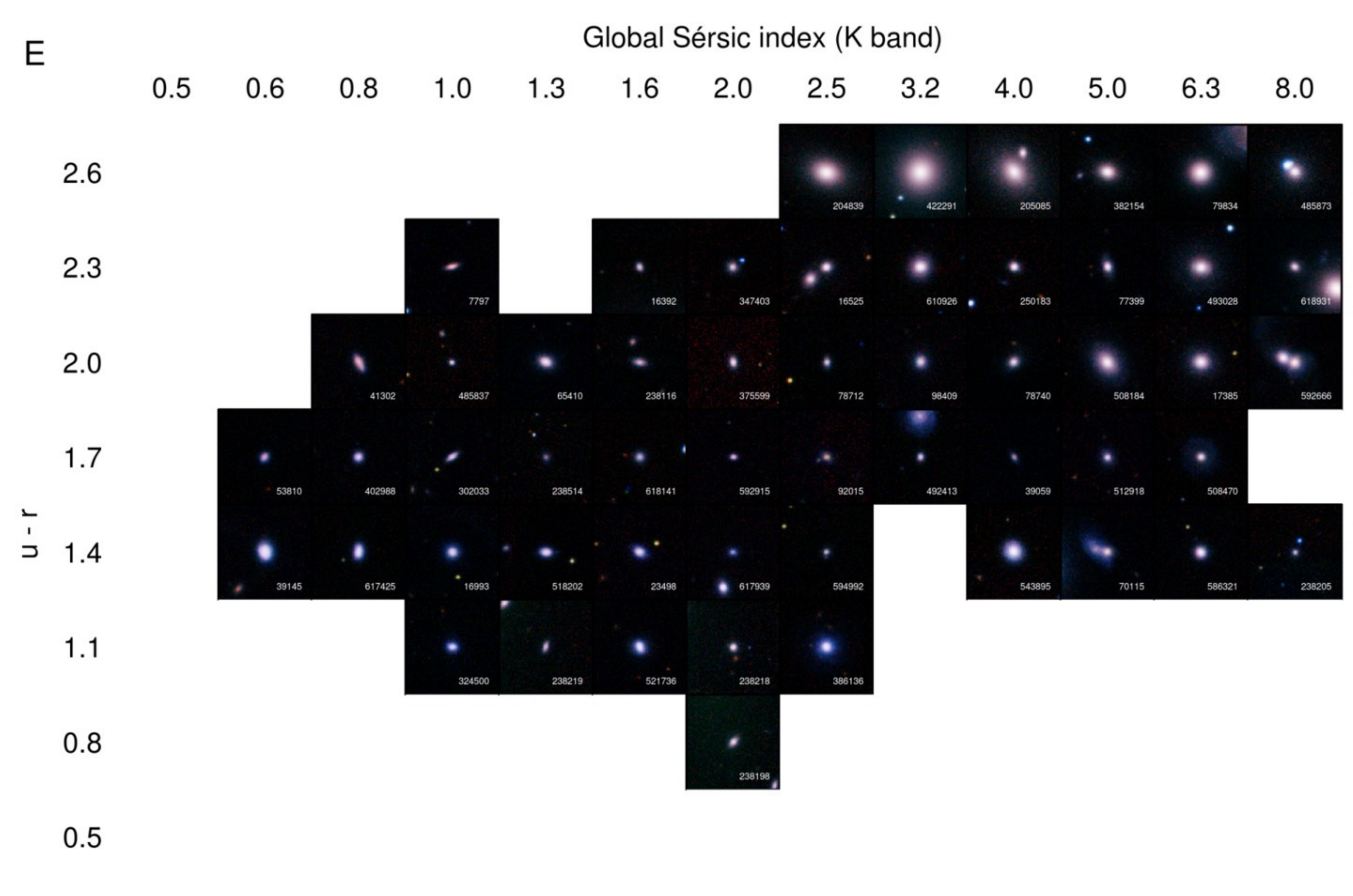}

\caption{\label{fig:eyeexam_E}As Figure \ref{fig:eyeexam_LBS}, but for ellipticals.}
\end{figure*}

\begin{figure*}
\includegraphics[width=0.9\textwidth]{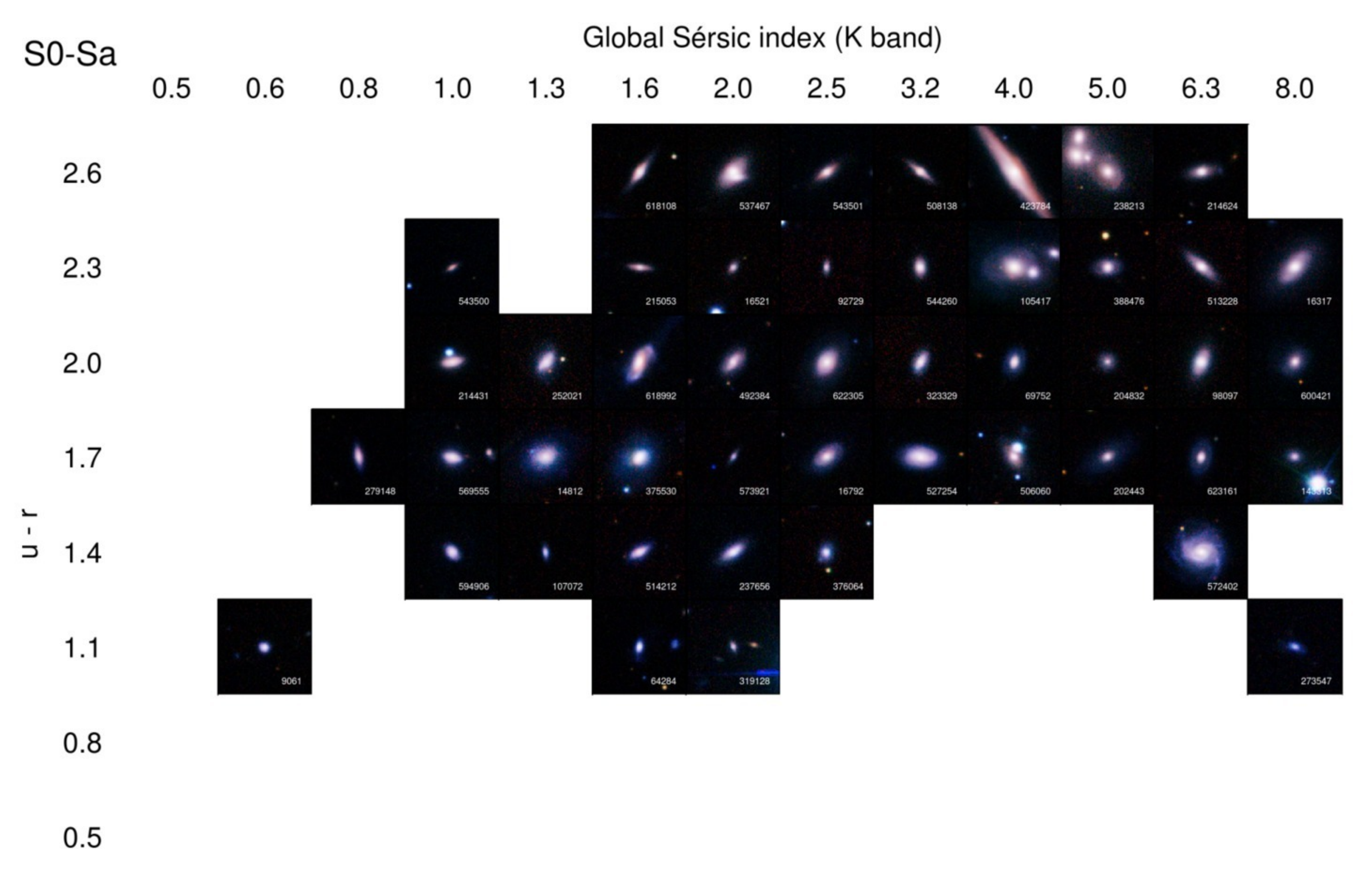}

\caption{\label{fig:eyeexam_S0a}As Figure \ref{fig:eyeexam_LBS}, but for
S0-Sa type galaxies.}
\end{figure*}

\begin{figure*}
\includegraphics[width=0.9\textwidth]{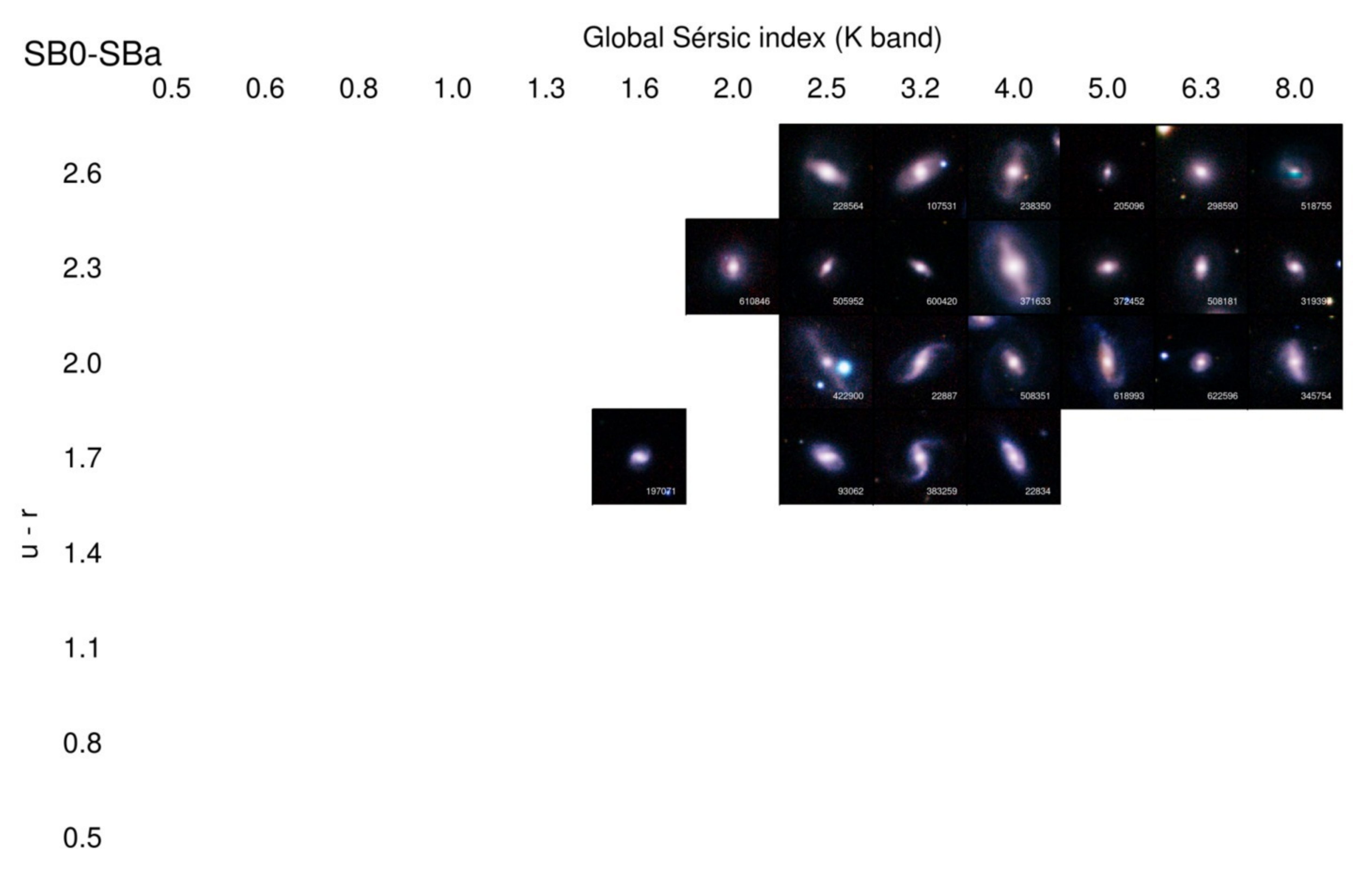}

\caption{\label{fig:eyeexam_SB0a}As Figure \ref{fig:eyeexam_LBS}, but for
SB0-SBa type galaxies.}
\end{figure*}

\begin{figure*}
\includegraphics[width=0.9\textwidth]{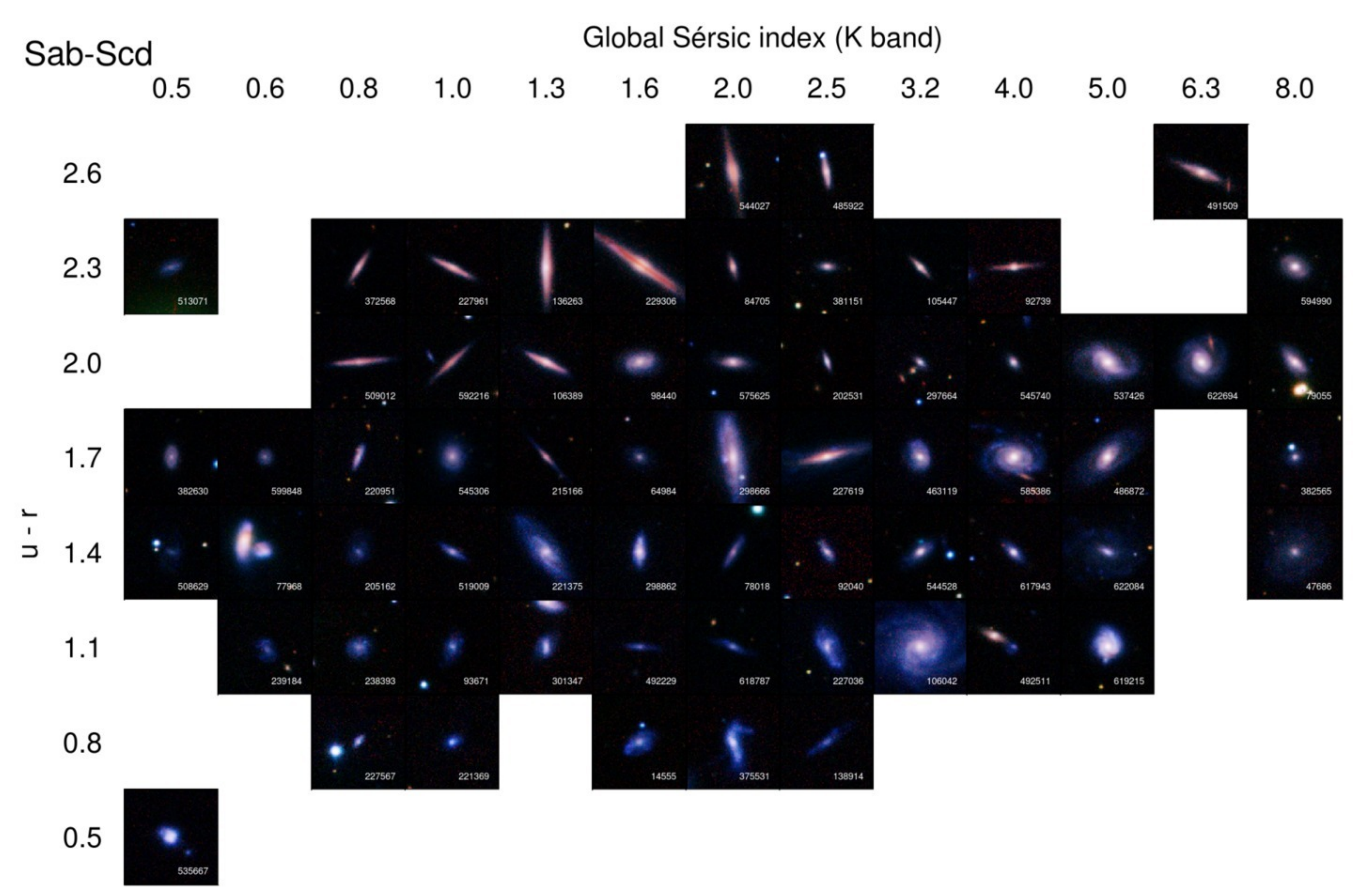}

\caption{\label{fig:eyeexam_Sbc}As Figure \ref{fig:eyeexam_LBS}, but for
Sab-Scd type galaxies.}
\end{figure*}

\begin{figure*}
\includegraphics[width=0.9\textwidth]{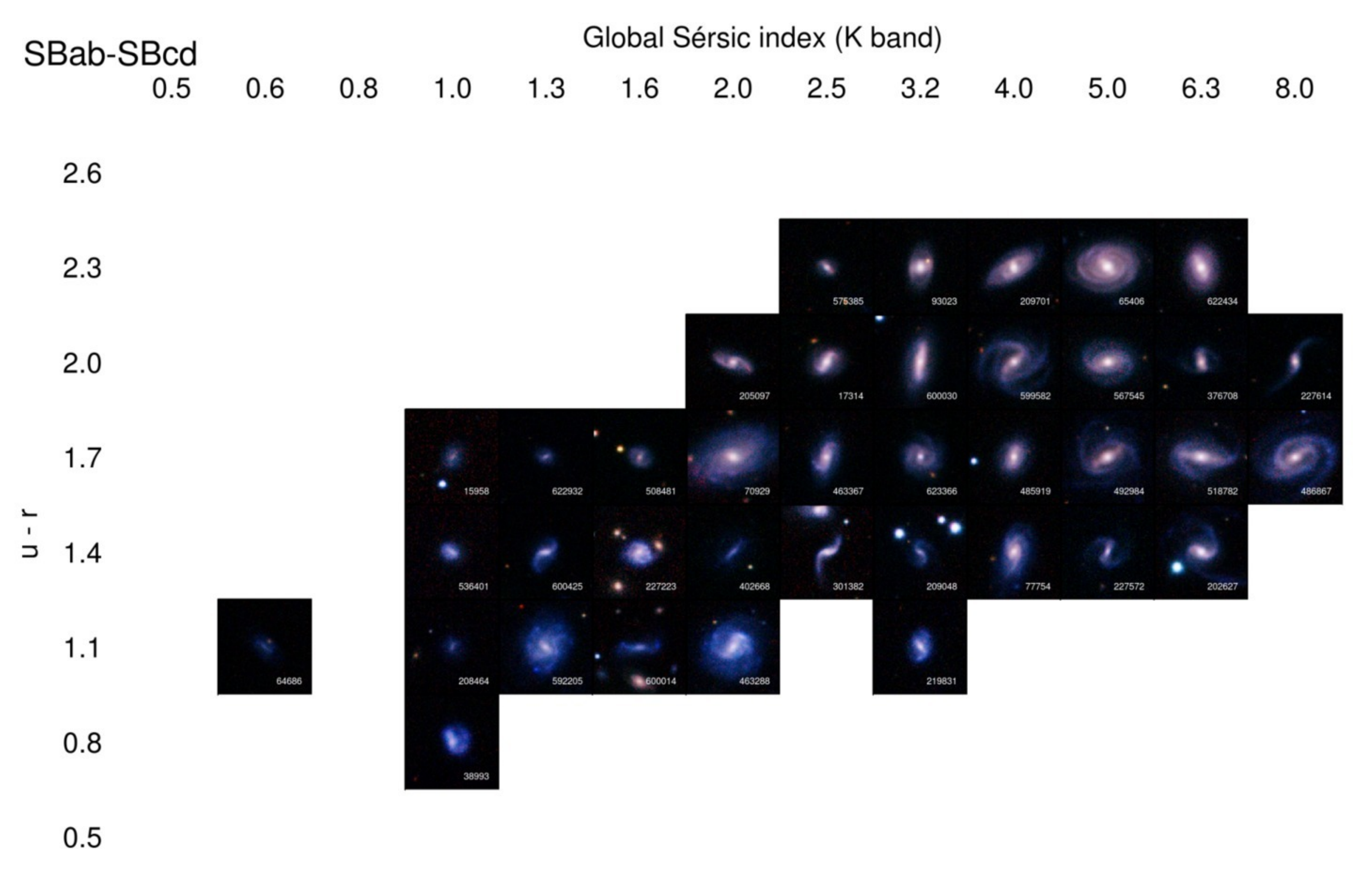}

\caption{\label{fig:eyeexam_SBbcd}As Figure \ref{fig:eyeexam_LBS}, but for
SBab-SBcd type galaxies.}
\end{figure*}

\begin{figure*}
\includegraphics[width=0.9\textwidth]{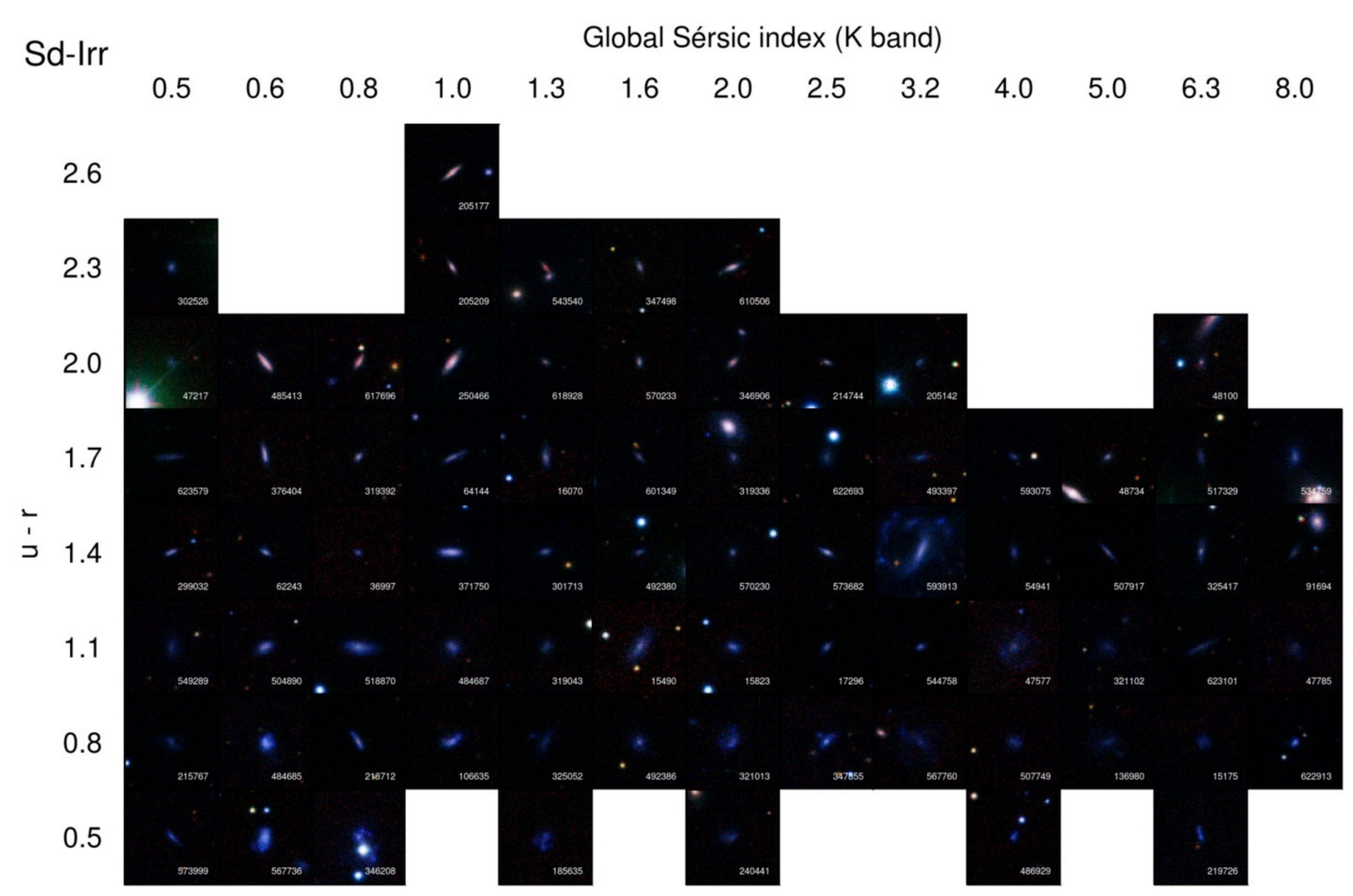}

\caption{\label{fig:eyeexam_Sd}As Figure \ref{fig:eyeexam_LBS}, but for Sd-Irr
type galaxies.}
\end{figure*}

Below we provide postage stamp examples of
each morphological type as defined in Figure \ref{fig:numbers}. These
types are: Little Blue Spheroids (LBS), Figure \ref{fig:eyeexam_LBS};
ellipticals, Figure \ref{fig:eyeexam_E}; lenticular/early-type spirals,
Figure \ref{fig:eyeexam_S0a}; barred lenticular/early-type spirals,
Figure \ref{fig:eyeexam_SB0a}; late-type spirals, Figure \ref{fig:eyeexam_Sbc};
barred late-type spirals, Figure \ref{fig:eyeexam_SBbcd}, and; disk-dominated
spirals, Figure \ref{fig:eyeexam_Sd}. Each figure is arranged according
to the global $K$ band Sérsic index and rest-frame $u-r$ colour
of the galaxy. Postage stamp images are created from RGB=$Hig$ input
data and are approximately $40''\times40''$ in size.

\label{lastpage}
\end{document}